\documentstyle[12pt]{article}
\begin{document}
\newdimen\theight
\newcommand{\ba}{\begin{array}}
\newcommand{\ea}{\end{array}}
\newcommand{\bea}{\begin{eqnarray}}
\newcommand{\eea}{\end{eqnarray}}

\def \Column{%
             \vadjust{\setbox0=\hbox{\sevenrm\quad\quad tcol}%
             \theight=\ht0
             \advance\theight by \dp0    \advance\theight by \lineskip
             \kern -\theight \vbox to \theight{\rightline{\rlap{\box0}}%
             \vss}%
             }}%

\catcode`\@=11
\def\qed{\ifhmode\unskip\nobreak\fi\ifmmode\ifinner\else\hskip5\p@\fi\fi
 \hbox{\hskip5\p@\vrule width4\p@ height6\p@ depth1.5\p@\hskip\p@}}
\catcode`@=12 

\def\cents{\hbox{\rm\rlap/c}}
\def\miss{\hbox{\vrule height2pt width 2pt depth0pt}}

\def\vvert{\Vert}                

\def\tcol#1{{\baselineskip=6pt \vcenter{#1}} \Column}

\def\dB{\hbox{{}}}                 
\def\mB#1{\hbox{$#1$}}             
\def\nB#1{\hbox{#1}}               
\newcommand{\be}{\begin{equation}}
\newcommand{\ee}{\end{equation}}

\begin{titlepage}
\begin{center}
\vspace*{2cm}

{\Large\bf Exact and quasi-exact solvability of two-dimensional
superintegrable quantum systems.

\vspace{0.4cm}
I. \, Euclidean space}

\vspace{0.7cm}

{\bf E.\ G.\ Kalnins}\\

{\sl Department of Mathematics and Statistics,
 University
of Waikato,}\\{\sl Hamilton, New Zealand.}\\

\vspace{0.2cm}
{\bf W.\ Miller, Jr.}

{\sl School of Mathematics, University of Minnesota,}\\
{\sl Minneapolis, Minnesota,
55455, U.S.A.}\\

\vspace{0.2cm}
{\bf G.\ S.\ Pogosyan}\\

{\sl Laboratory of Theoretical Physics, Joint Institute for
Nuclear Research,}\\
{\sl Dubna, Moscow Region 141980, Russia}
\\
[2mm]
and
\\
[2mm]
{\sl Centro de Ciencias F\'{\i}sicas Universidad Nacional Aut\'onoma
de M\'exico, \\
Apartado Postal 48--3, 62251 Cuernavaca, Morelos,
M\'exico}

\end{center}

\vspace{1cm}

\vspace{1cm}
\begin{abstract}
In this article we show that  separation of variables for
second-order superintegrable systems in two-dimensional Euclidean
space generates both  exactly solvable (ES) and quasi-exactly solvable
(QES) problems in quantum mechanics. In this article we propose the
another definition of ES and QES. The quantum mechanical problem is
called ES if the solution of Schroedinger equation, can be expressed
in terms of hypergeometrical functions $_mF_n$ and is QES if the
Schroedinger equation admit polynomial solutions with the coefficients
satisfying the three-term or more higher order of recurrence relations.

\end{abstract}
\vspace{2cm}
\today
\end{titlepage}

\section{Introduction}

It is well known that $N$-dimensional nonrelativistic quantum
systems described by the Hamiltonian
\bea
\label{HAM1}
{\cal H} =
- \frac{1}{2} \sum^{N}_{i=1}
\frac{\partial^2}{\partial x_i^2} + V(x_1, x_2,..., x_N)
\eea
are integrable if there exist $N$ linearly independent and global
integrals of motion ${\cal I}_{\ell}$,
$\ell=0,1,..N-1$ and ${\cal I}_{0}={\cal H}$, commuting with the
Hamiltonian (\ref{HAM1}) and with each other
\bea
\label{HAM2}
[{\cal I}_{\ell}, {\cal H}] = 0,
\qquad
[{\cal I}_{\ell}, {\cal I}_{j}] = 0,
\qquad
\ell, j = 1,2,...N-1.
\eea
This particular class of integrable systems is called superintegrable
(this term was introduced first time by S.Rauch-Wojciechowski in
\cite{SWOJ}) if it is integrable and, in addition to this,
possesses more integrals of motion than degrees of freedom.
The additional integrals ${\cal L}_{k}$, commute with Hamiltonian
\bea
\label{HAM3}
[{\cal L}_{k}, {\cal H}] = 0,
\qquad
k = 1,2,...N,
\eea
but not necessarily with each other. If the number $D$ of linearly
independent integrals takes the value $D=2N-1$ (N the number of
degrees of freedom) then the system is called {\it maximally}
superintegrable \cite{EVANS} and it is called {\it minimally}
superintegrable if it has $D=2N-2$ integrals of motions.
Three examples of this kind  have been well-known a long time, namely the Kepler-Coulomb problem, the isotropic harmonic
oscillator, and the nonisotropic oscillator with commensurable
frequencies.

The existence of additional integrals of motion for these systems
leads to many interesting properties unlike standard integrable
systems. In particular, in quantum mechanics, there is phenomenon
of {\it accidental degeneracy} when all energy eigenvalues are
multiply degenerate. This property is intimately related to the
existence of a dynamical symmetry group, a so-called {\it hidden
symmetry group}, which contains properly the  geometrical symmetry
group describing this system. For instance, the dynamical group
for the  hydrogen atom is O(4)  for the discrete spectrum \cite{FOCK1},
and the Lorentz group O(3,1) for the continuous spectrum \cite{BANDER}.
For the  isotropic harmonic oscillator it is  SU(3) \cite{BAKER}.

In classical mechanics the additional integrals of motion have the
consequence that in the case of  superintegrable systems in two
dimensions and maximally superintegrable systems in three dimensions
all finite trajectories are found to be periodic; in the case of
minimally super-integrable systems in three dimensions all finite
trajectories are found to be quasi-periodic \cite{KIBLER1}.

One of the most important properties for superintegrable systems is multiseparability, the
separation of variables for the Hamilton-Jacobi and Schr\"odinger equations
in more than one orthogonal coordinate system \cite{EIS,MIL,ERNIE}.
For instance, the isotropic harmonic oscillator in three dimensions is
separable in eight coordinate systems, namely in Cartesian, spherical,
circular polar, circular elliptic, conical, oblate spheroidal, prolate
spheroidal, and ellipsoidal coordinates.
The Kepler-Coulomb potential is separable in four coordinate systems,
namely in conical, spherical parabolic, and prolate spheroidal
coordinates.

A systematic search for such systems in two- and three-dimensional
Euclidean space was started in the pioneering work of Smorodinsky
and Winternitz with collaborators in \cite{FMSUW,FSUW,MSVW} and was continued
in \cite{EVANS}. Particularly, in \cite{FSUW} it was shown that in
two-dimensional Euclidean space there exist four superintegrable potentials
(see Table 1.), three of them could be considered as the singular
generalization of Kepler-Coulomb, circular oscillator and anisotropic
oscillator systems. These results were extended for  two- and
thee-dimensional spaces with constant curvature (both positive and
negative) \cite{GPS1}, and on the complex two-dimensional
sphere and Euclidean space \cite{MKP1,MKP2,MKP3,MKP4,GPS0}.
The same program is continuing nowadays both in  spaces with constant
and nonconstant curvature
\cite{KMP-96,KIBHAK,KWMPOG,KMP-HYP,RANAD1,KMPW1,KRESS1,MKW1,HER1,Gonera1,MKW2,GRAVEL}.

The wide  applicability of superintegrable systems, both in physics
and mathematical physics has stimulated  further investigations.
In the last fifteen years  superintegrable systems (in spaces of
constant curvature including flat Euclidean space) became a subject
of investigation from  many point of view:
in \cite{GPS1,GPS0,BIJ1,BIJ2} via the path integral
approach, in \cite{KMP-96,KWMPOG,KMP-HYP} by solving the
Schr\"odinger equation
with the help of the Niven ansatz \cite{WITTEKER}, in
\cite{HIGGS,LEEMON,ZEDAN1,VINET,DASKALO1,DASKALO2}
from the purely algebraic approach.
As has been shown by a number of authors, almost all
superintegrable systems (excluding the pure $N$-dimensional
Kepler-Coulomb and harmonic oscillator potentials) generate an
algebraic structure which may be considered as a nonlinear extension
of the Lie algebra (in classical
mechanics Poisson algebras), namely a quadratic algebra. The general
form of quadratic algebras, which are encountered in the case of
two-dimensional quantum superintegrable systems has been
investigated by Daskaloyannis \cite{DASKALO2}.

In spite of all the  above listed characteristics, superintegrable systems
would not be so useful, except for the property
of {\bf exact solvability} of the all known superintegrable systems
(at least in case of the second kind of superintegrability).
More precisely this means that after any separation of variables each
of the separated ordinary differential equations admits an  exact solution.
This question, again quite recently has been discussed in the
literature \cite{WINTER1,WINTER2}. As mentioned in these papers
the term exact solvability is defined quite differently by different authors.
\footnote
{In \cite{WINTER1,WINTER2} (see also the recent paper
\cite{KAMRAN1}) we read that ``an exactly solvable quantum
mechanical system can be characterized by the fact that in its solution
space one can indicate explicitly an infinite flag of functional linear
spaces, which is preserved by the Hamiltonian" or the
``Hamiltonian is exactly solvable if its spectrum can be calculated
algebraically".}
Indeed, in spite of an ``intuitive" understanding of the term exactly
solvable, no universal definition exists up to now.

On the other hand  there are limiting cases of well-known one-dimensional
exactly solvable systems, namely the harmonic oscillator and Coulomb problems
with ${\gamma}/{x^2}$ ($\gamma > -1/4$) interaction, Morse potential,
trigonometric and modified P\"oschl-Teller potentials, trigonometric
and hyperbolic Manning-Rosen potentials \cite{FLUGGE,LANGE}, and the
Natanson potential \cite{NATA}.
All these potentials have the general property that the Schr\"odinger
spectral problem for bound state (or continuous state) has an
explicit formula for the {\it whole} energy spectrum, and the
eigenfunctions (up to the asymptotic ansatz or gauge
transformations \cite{WINTER1,WINTER2})
are of hypergeometric type $_1F_1$, $_2F_1$. For the
bound states we have solutions in term of classical polynomials
\cite{BE} whereas for continuous states just infinite series.
Moreover, hypergeometric functions describe both the continuous
quantum systems  as well as the finite systems and appear also
as solutions of related difference equations, for instance, the
finite one- and two-dimensional oscillator expressed in terms
of discrete variables polynomials: Krawchuk, Meixner and Hahn
\cite{ATAK}.

Thus, we propose another definition of exact-solvability:
a {\it quantum mechanical system is called exactly solvable if the
solutions of Schr\"odinger equation, can be expressed in terms of
hypergeometric functions $_mF_n$}. (Basically, we are requiring that
the coefficients in power series expansions of the solutions satisfy
two-term recurrence relations, rather than recurrence relations of
higher order.)
It is obvious, that an $N$-dimensional Schr\"odinger equation is
exactly-solvable  if it is separable in some coordinate system
and each of the separated equations is exactly-solvable.
Further,
{\it We say  that a superintegrable system is exactly solvable
if is exactly solvable  in at least one system of coordinates}.

At first sight, such a definition of exactly-solvable problems may
seem too narrow, but  it leads us to distinguish two kind of models: 1) those which
is possible to study analytically and 2) those which admit just
numerical solutions
(even if they admit  polynomial solutions and their energy spectra can
be calculated by solving an algebraic equation or system of equations).

The process of separation of variables in the $N$- dimensional Schr\"odinger
equation leads to ordinary differential equations having as solutions
many special functions of mathematical physics. A further complication of
the separated equations involves the $N$ separation constants.
In general we have a multiparameter eigenvalue problem \cite{SLEEMAN}.
It is possible to
distinguish  three different cases, namely when there is complete,
partial or non-separability of the separation constants. It is obviously
that in the case of complete separability (of separation constants) the
initial $N$-dimensional Schr\"odinger equation splits into $N$
independent second order differential equations,
each involving a single separation parameter.
This situation occurs, for instance,
in the case of separation of variables in the
Helmholtz (free Schr\"odinger equation, which is also superintegrable)
or the Schr\"odinger equation for the  harmonic oscillator in Cartesian
coordinates. The second "extremal" case, when  complete
non-separability exits, is realized,  in separation
of variables for the same problems but in ellipsoidal coordinates.
In the last case the each separated second-order differential equation
contains  at once all separation constants (usually depending from
dimensional or non-dimensional parameters) \cite{MIL,ERNIE},
for which the simultaneous quantization becomes nontrivial.

The standard method of solution of  a second
order ordinary differential equation, obtained after separation of variables
in $N$ - dimensional Schr\"odinger equations,  involves (after taking
into account the asymptotic ansatz) expansions around one of the
singular points of the differential equation (standard power-series method
\cite{INCE}, or the so-called Hill-determinant method \cite{WATSON}).
After that the problem reduces to solution of the recurrence relations for the expansion coefficients.
If the coefficients obey a two-term recurrence relation,  then the
corresponding solution will be written in closed or analytic form or
in terms of hypergeometric functions and we have an  {\bf exactly-solvable}
problem.
Such situations occur when separation of variables for
superintegrable systems is possible
in sub-group type coordinate  (spherical, cylindrical and Cartesian)
\cite{POGWINT} and often in parabolic type coordinates.
This method also powerfull when separation of variables possible in
non-subgroup systems of coordinates as spheroidal or elliptic. In this
case we arrive to high-order recurrence relation, the subsequent
analysis of which, allow to investigate the behavior of the solution
and to answer on the question whether exist the finite solution.

Actually there is another general approach for solving the Schr\"odinger
equation by exploring the Niven - type ansatz \cite{WITTEKER} and
essentially is based on existing of finite solutions. According to
this methods the complete solution can be constrated without direct
separation of variables and computed in terms of zeros of corresponding
polynomial. This method have been using in papers
\cite{KMP-96,KWMPOG,KMP-HYP} at the investigation of two- and
three-dimensional superintegrable systems in Euclidean and curve
spaces.

Consider now the problem of motion in the plane for a charged particle
with two fixed Coulomb centers with coordinates ($\pm D/2, 0$)
(so-called plane two center problem)
\bea
\label{HAM12}
V(x,y) =
- \frac{\alpha_1}{\sqrt{y^2 + (x+D/2)^2}}
- \frac{\alpha_2}{\sqrt{y^2 + (x-D/2)^2}}
\eea
This system is not superintegrable and separation of variables is
possible only in two-dimensional elliptic coordinates
(see eq. (\ref{SCH-V2-EL70})).
Upon the substitution $\psi (\nu, \mu; D^2) = X(\nu; D^2) Y(\mu; D^2)$
and  the separation constant $A(D)$, the Schr\"odinger
equation splits into a {\bf system} of two ordinary differential equations
\bea
\frac{d^2 X}{d\nu^2}
&+&
\left[ \frac{D^2 E}{2} \cosh^2\nu + D(\alpha_1 + \alpha_2) \cosh\nu
+ A(D)\right]X = 0,
\label{SCH-V0-EL1}
\\[3mm]
\frac{d^2 Y}{d\mu^2}
&-&
\left[\frac{D^2 E}{2}\cos^2\mu + D(\alpha_1 - \alpha_2)\cos\mu
+ A(D)\right] Y = 0.
\label{SCH-V0-EL2}
\eea
Both  equations (\ref{SCH-V0-EL1})-(\ref{SCH-V0-EL2}) belong to
the class of {\bf non-exactly solvable} problem. In generally
the polynomial solutions do not exist even for the case of discrete spectrum
$E<0$\footnote{To be completely correct let us note that the polynomil
solution exist only for special values of parameters $\alpha_1, \alpha_2$
and $R$.},
and each of the wave functions $X(\nu; D^2)$ and $Y(\mu; D^2)$
is expressed as an  infinite series with a three-term recurrence relation.

Let us now  put $\alpha_2=0$. Then the potential (\ref{HAM12}) transforms
to the ordinary two-dimensional (2d) hydrogen atom problem,
which is well-known
as a superintegrable system \cite{T-4,DULOCK,ENGLEF} with dynamical
symmetry group $SO(3)$, and admits separation of variables in three
systems of coordinates: polar, parabolic and elliptic.
In this case we can see that the separation equations
(\ref{SCH-V0-EL1}) and (\ref{SCH-V0-EL2}), namely
\bea
\frac{d^2 X}{d\nu^2}
&+&
\left[ \frac{D^2 E}{2} \cosh^2\nu + D \alpha_1 \cosh\nu
+ A(D)\right]X = 0,
\label{SCH-V0-EL3}
\\[3mm]
\frac{d^2 Y}{d\mu^2}
&-&
\left[\frac{D^2 E}{2}\cos^2\mu + D \alpha_1 \cos\mu
+ A(D)\right] Y = 0.
\label{SCH-V0-EL4}
\eea
transform into each other by the change $\mu \leftrightarrow i\nu$.
Thus separation of variables in elliptic coordinates for the
2d hydrogen atom gives two functionally identical
one-dimensional Schr\"odinger type equations with two parameters:
coupling constant $E$ and energy $A(D)$ (correspondingly energy and
separation constant for 2d), but one defined on the real and the
other on the imaginary axis.
In other word, instead the systems of differential
equations (\ref{SCH-V0-EL3})-(\ref{SCH-V0-EL4}), the task reduces to
solving only to the one of equations (\ref{SCH-V0-EL3}) or
(\ref{SCH-V0-EL4}) for what the "domain of difinition" is complex plane.
The requirement of finiteness for the wave functions in the complex
plane permits {\bf only} polynomial solutions (see for details
\cite{MPSTAN1}).
As result we obtain {\it simultaneously} quantization of the energy
spectrum
\bea
E_n = - \frac{\alpha_1^2}{2(n+ 1/2)^2},
\qquad n=0,1,2,...
\label{SCH-V0-EL5}
\eea
and the elliptic separation constant $A_s(D)$ where $s=0,1,2,...n$
(as a solution of an $n$th-degree algebraic equation).
The polynomial solution difine by the help of
finite series with the three-term recurrent relations for coefficients.
They cannot be considered as exactly-solvable and maybe investigated
only numerically.
A similar situation occurs, for instance, in the case of the two-center
problem in three-dimensional Euclidean space (the so-called prolate
spheroidal radial and angular Coulomb wave functions) \cite{PONOM}
and three-dimensional sphere (Heun wave functions) \cite{BOT}, where
after eliminating one of the Coulomb centers the problems have reduced
to superintegrable systems admitting only polynomial solutions.
The presented above (and many others) examples let us to claim about
deep connection of the notion of superintegrability and existing of
polynomial solutions of the corresponding Schr\"odinger equation.

At the other hand side the each of equations (\ref{SCH-V0-EL3}) or
(\ref{SCH-V0-EL4}) have the form of one dimensional Schr\"odinger
equation with the parameter $E$ and for eigenvalue $A(D)$, and
could be separately considered in the regions $\mu\in [0, 2\pi]$
or $\nu\in [0, \infty)$ correspondingly. Then for the arbitrary
values of constant $E$ \footnote{For example when $E_n=0$ ($n\to \infty$)
the equations (\ref{SCH-V0-EL3}) and (\ref{SCH-V0-EL4}) transforms
to periodic and modifying Matie equations, which are non-exactly
solvable.}
the solutions of eqs. (\ref{SCH-V0-EL3}) or (\ref{SCH-V0-EL4})
expressed via infinite series and only on the "energy surface"
corresponding of 2d hydrogen atom (\ref{SCH-V0-EL5})
the solutions split onto polynomial and nonpolynomial sectors
(each of these sectors are non complete) and for fixed number $n$,
only {\it part} of eigenvalue $A_s(D)$, ($s=0,1,2...n$) are possible
to calculate as nth-degree algebraic equation.
We can say that the eqs. (\ref{SCH-V0-EL3})
and (\ref{SCH-V0-EL4}) "remember" for its polynomial solutions.
It is obviously that the spectrum of $A_s(D)$, ($s=0,1,2...n$)
and solutions in polynomial form of the each of equations
(\ref{SCH-V0-EL3})-(\ref{SCH-V0-EL4}) coinside with the eigenvalue of
separation constant and the wave function after the reduction to
one of the regions $\mu\in [0, 2\pi]$ or $\nu\in [0, \infty)$]
for 2d hydrogen atom.

This phenomena have been intensively discussed in literature in the
late 1980's and was determined as {\it quasi-exactly solvability}
(this term first time introduced by Turbiner and Ushveridze in
\cite{PREP}) and such models called as {\it quasi-exactly solvable}
systems \cite{TURBO,USHV1,SHIF1} (see also \cite{USH} and references
therein).
The crucial example stimulated the investigation of quasi-exactly
solvable systems is the hamiltonian (\ref{HAM1}) with the anharmonic
potential
\bea
V (x) =
\frac{1}{2} \omega^2 x^6 + 2\beta \omega^2 x^4 +
(2\beta^2 \omega^2 - 2\delta\omega - \lambda)x^2
+ 2 \frac{(\delta-\frac14)(\delta-\frac34)}{x^2},
\label{SCH-V0-ANHAR0}
\eea
where $\omega$, $\beta$, $\delta > 1/2$ and $\lambda$ are the constants.
As it was notice of many authors \cite{BISWAS,TURBIN-USH,BENDER},
this systems only for special values of constant
$\lambda = \omega(2n + 1)$, ($n=0,1,2...$)
admit partial polynomial solutions
\bea
\Psi (x) \approx
x^{2\delta - \frac12} \,
{\rm e}^{- \frac{\omega}{4} x^4 - \beta \omega x^2} \,
P_n (x^2).
\label{SCH-POLYNOM1}
\eea
There are two different approaches in investigation of quasi-exactly
solvable systems. In the base of algebraic approach formulated by
Turbiner in \cite{TURBO} the fact of quasi-exactly solvabilily was
explain in term of "hidden symmetry algebra" $sl(2,R)$
\footnote{Really this is not hidden dynamical symmetry in usual sence
because the hamiltonian (\ref{SCH-V0-ANHAR1}) belong to the
enveloping algebra but not a Casimir.}.
More precisely it means following:
the one-dimensional Hamiltonian (\ref{HAM1}) after suitable changes
of variable $z = \xi(x)$ and "gauge transformation"
$H= e^{-\alpha(z)}{\cal H} e^{\alpha(z)}$ can be written in form
\bea
H =
\sum_{a,b = 0, \pm} C_{ab} J_a J_b
+
\sum_{a = 0, \pm} C_{a} J_a
\label{SCH-V0-ANHAR1}
\eea
where the first-order differential operators $\{J_{\pm}, J_0\}$
span to the  commutation relations for algebra $sl(2,R)$
\cite{TURBO}.

The mentioned above analysis for the 2d hydrogen atom have shown,
that in spite of the elegance of algebraic approach, the phenomena
of quasi-exactly solvability have really more deeper roots that
it can be explain via "one-dimensional" model (\ref{SCH-V0-ANHAR1}).
The other examples are hydrogen atom and oscillator problems on two
and three-dimensional spheres \cite{KMP-96,POG-HAKOB1} and
two-dimensional hyperboloids \cite{KMP-HYP}, which generate not
only hyperbolic and trigometric but elliptic type of quasi-exactly
solvable systems (se also oldest articles
\cite{AKPS,MPSTAN1,MPSTAN,DAVTYAN1}). We can also mention Lam'e
polynomials. They comes from separation of variables for the
Helmholtz (also superintegrable!) or Schr\"odinger equation in
elliptic coordinates on the two-dimensional sphere.
As it have been also determined in \cite{VINET} (without
showing of the mechanism of this phenomena) some of the
quasi-exactly solvable systems can be obtained through dimensional
reduction from the two and three-dimensional superintegrable
models with quadratic invariants (second-order superintegrability).

The second approach, known as analitic, was formulated by
Uschveridze (see for example \cite{USHV1,SHIF1,USH}) and really
represent a one-dimensional reduction of the Niven-Stilties method
for solving of many-parametric differential equation as generalized
Lame equation (or ellipsoidal equation) \cite{WITTEKER}.
The solution in this method are presented in the term of zeros of
polynomials $P_n(x^2)$. Then the wave function (\ref{SCH-POLYNOM1})
can be rewritten in the form
\bea
\Psi (x) \approx
x^{2\delta - \frac12} \,
{\rm e}^{- \frac{\omega}{4} x^4 - \beta \omega x^2} \,
\Pi_{i=0}^{n} (x^2 - \xi_i),
\label{SCH-POLYNOM2}
\eea
where the numbers $(\xi_1, \xi_2, ... \xi_n$) satisfying the systems
of $n$ algebraic equations (see section 2.3).
Accordingly to the oscillation theorem the number of zeros in
physical interval $\xi_i \in [0, \infty)$ enumerate the ground state
and first $n$ - excitations, describing in terms of all zeros
(complete solutions of the systems of algebraic equations and
including non physical section $\xi_i \in (-\infty, 0]$)
as
\bea
E = 4\delta \left[ \beta \omega + \sum_{i=1}^{n} \frac{1}{\xi_i}
\right].
\label{SCH-POLYNOM3}
\eea
The two nutural questions appering in such approach: what is the
physical meaning of the negative zeros $\xi_i$, and why in the correct
formula of energy spectrum (\ref{SCH-POLYNOM3}) are participating all
$n$ zeros of the polynomial $P_n (x^2)$?

With this article we begin the new investigation of all known two-
and three-dimensional second order superintegrable systems on curved
spaces (Euclidean, sphere, hyperboliod and pseudo-euclidean) base on
the separation of variables and direct solutions of the Schr\"odinger
equation. We pay special attention to non-subgroup type coordinates
and prove the existence the polynomial solution for this systems,
which we found surprisingly that never have been done before.

We also demonstrate that the fact of quasi-exactly solvability are
directly related with the multiseparability of the second-order
superintegrable systems, from one hand side, and with presence of
polynomial solution in this systems on the other hand side.

The present work devoted only two (singular anisotropic and
singular circular oscillators) from four famous superintegrable
systems in two-dimensional Euclidean space (see first two
potentials in Table 1.).
The next two systems maybe transformed (only for the discrete spectrum!)
to singular circular oscillator (for $V_3$) or ordinary shifted
oscillator (for $V_4$) systems by the help of Levi-Civita mapping
\cite{KS}.

\vspace{0.5cm}
\hfuzz=25.0pt
\medskip
\begin{center}
{{\bf Table 1:} Superintegrale potentials in two-dimensional
Euclidean space.\hfill}
\end{center}
\begin{eqnarray}
\begin{array}{l}
\vbox{\offinterlineskip
\hrule
\halign{&\vrule#&\strut\quad\hfil#\hfill\cr
height2pt&\omit&&\omit&\cr
&Potential $V(x,y)$
  &&Coordinate                                              &\cr
& &&\ System                                                &\cr
height2pt&\omit&&\omit&&\omit&\cr
\noalign{\hrule}
\noalign{\hrule}
height2pt&\omit&&\omit&&\omit&\cr
&$\displaystyle
  V_1=\frac{1}{2}\omega^2(4x^2+y^2)+k_1x
     + \frac{1}{2}{{k_2^2-\frac{1}{4}}\over y^2}$
  &&${\hbox{Cartesian}}$              &\cr
& &&Parabolic                                   &\cr
height2pt&\omit&&\omit&\cr
\noalign{\hrule}
height2pt&\omit&&\omit&\cr
&$\displaystyle
  V_2=\frac{1}{2}\omega^2(x^2+y^2)+ \frac{1}{2}
    \Bigg({{k_1^2-\frac{1}{4}}\over x^2}+
    {{k_2^2-\frac{1}{4}}\over y^2}\Bigg)$
  &&${\hbox{Cartesian}}$              &\cr
& &&${\hbox{Polar}}$                  &\cr
& &&Elliptic                                    &\cr
height2pt&\omit&&\omit&\cr
\noalign{\hrule}
height2pt&\omit&&\omit&\cr
&$\displaystyle
  V_3=-{\alpha\over\sqrt{x^2+y^2}}+ \frac{1}{4}
    {1\over\sqrt{x^2+y^2}}\Bigg({{k_1^2-\frac{1}{4}}\over\sqrt{x^2+y^2}+x}
         +{{k_2^2-\frac{1}{4}}\over\sqrt{x^2+y^2}-x}\Bigg)$
  &&${\hbox{Polar}}$                  &\cr
& &&Elliptic II                                 &\cr
& &&${\hbox{Parabolic}}$              &\cr
height2pt&\omit&&\omit&\cr
\noalign{\hrule}
height2pt&\omit&&\omit&\cr
&$\displaystyle
  V_4= -{\alpha\over\sqrt{x^2+y^2}}+ \frac{1}{4}
   \frac{1}{\sqrt{x^2+y^2}}
   \left(\beta_1\sqrt{\sqrt{x^2+y^2}+x} + \beta_2
   \sqrt{\sqrt{x^2+y^2}-x}\right)$
     &&${\hbox{Mutually}}$               &\cr
& &&${\hbox{\ Parabolic}}$            &\cr
height2pt&\omit&&\omit&\cr}\hrule}
\end{array}\nonumber
\end{eqnarray}
\hfuzz=7.0pt

\newpage
\section{Singular anisotropic oscillator}

Let us first consider the potential \, ($k_2 > 0$)
\be
\label{V11}
V_1(x,y)= {1\over 2}\omega^2(4x^2+y^2) + k_1 x +
{k^2_2 - {1\over 4}\over 2y^2}
\ee
(second potential on the table 1.),
which we will call the {\it singular anisotropic oscillator}.
The Schr\"odinger equation has the form
\be
\label{V1-EQ0}
\left({\partial^2 \over \partial x^2} + {\partial^2\over \partial y^2}
\right)\Psi
+
\left[2E - \omega^2 (4x^2+y^2) - 2k_1 x -
{k^2_2 - {1\over 4}\over y^2}\right]\Psi = 0.
\ee
For $k_2 > 1/2$ the singular term at $y=0$ is repulsive and the motion
takes place only on one of the half planes ($-\infty < x < \infty, \, y >0$)
or ($-\infty < x < \infty, \, y <0$), whereas for $0 < k_2 < 1/2$ in
whole plane ($x, y$). There are two coordinate systems of relevance
here: {\it Cartesian and parabolic coordinates}.

\subsection{Cartesian bases}

Separation of variables for the eq. (\ref{V1-EQ0}) in Cartesian
coordinates leads to the two {\it independent} one-dimensional
Schr\"odinger equations
\bea
\label{V1-SHEQ0}
\frac{d^2 \psi_1}{dx^2}
&+&
(2\lambda_1 - 4 \omega^2 x^2 - 2k_1 x )\psi_1 = 0.
\\[2mm]
\label{V1-SHEQ1}
\frac{d^2 \psi_2}{dy^2}
&+&
\left(2\lambda_2 - \omega^2 y^2 -
\frac{k^2_2 - \frac14}{y^2}\right)\psi_2 = 0.
\eea
where
\bea
\label{V1-SHEQ00}
\Psi(x,y; k_1, \pm k_2) = \psi_1(x; k_1)\psi_2(y;\pm k_2)
\eea
and $\lambda_1$, $\lambda_2$ are {\it two Cartesian separation constant}
with $\lambda_1 + \lambda_2 = E$.

The equation (\ref{V1-SHEQ1}) represents the well-known linear singular
oscillator system studied in detail (see for instance the books
\cite{GOLDMAN,LANDAU3} and articles \cite{FSUW,KIBHAK,HALL}).  It is
an exactly-solvable problem and has been used in many applications,
for example as a model in $N$ - body problems \cite{CALOGERO1}, or fractional
statistics and anyons \cite{ISAKOV,TERANT}. The complete set of
orthonormalized eigenfunctions, (on $1/2$) in the interval $0 < y < \infty$
of eq. (\ref{V1-SHEQ1}),  can be express in terms of finite confluent hypergeometric
series or Laguerre polynomials
\bea
\label{V1-SHEQ2}
\psi_{n_2} (y;\pm k_2) =
\sqrt{\frac{2\omega^{(1 \pm k_2)} n_2!}{\Gamma(n_2 \pm k_2+1)}}
\, y^{\frac{1}{2} \pm k_2} \, e^{-{1\over 2}\omega y^2}
\,
L^{\pm k_2}_{n_2}(\omega y^2)
\eea
and $\lambda_2 = \omega(2n_2+1 \pm k_2)$. We assume that the positive
sign at the $k_2$ has to be taken if $k_2>{1\over 2}$ and both the
positive and the negative sign must be taken if $0<k_2<{1\over 2}$,
so that the polynomials have finite norm. Let us also note that unlike
the potential (\ref{V11}) the wave function is not invariant under the
replacement $k_2 \to - k_2$ and splits into two families of solutions that transform to
one another under this change.

The second equation (\ref{V1-SHEQ0}) easily transforms to the ordinary
one-dimensional oscillator problem. In terms of Hermite polynomials
the orthonormal solution (in region $-\infty < x < \infty$) is
\bea
\label{V1-SHEQ3}
\psi_{n_1} (x; k_1)
=
\left({2\omega \over \pi }\right)^{1/4} \,
\frac{e^{-\omega z^2}}{\sqrt {2^{n_1}{n_1}!}}\,
H_{n_1}(\sqrt {2\omega }z),
\qquad\quad
z= x + \frac{k_1}{4\omega^2},
\eea
and where $\lambda_1 = \omega (2n_1+1) - \frac{k_1^2}{8\omega^2}$.
Thus the complete energy spectrum is
\be
\label{V2-E1}
E = \lambda_1 + \lambda_2 =
\omega [2n + 2 \pm k_2] - \frac{k_1^2}{8\omega^2},
\qquad n=n_1+n_2 = 0,1,2,...
\ee
and the degree of degeneracy for fixed principal quantum number $n$ is
$(n+1)$. Finally note that the separation of variables in Cartesian
coordinates leads to  two {\bf exactly solvable} one-dimensional
Schr\"odinger equations and the complete wave function may be constructed
with the help of formulas (\ref{V1-SHEQ2}), (\ref{V1-SHEQ3}) and
(\ref{V1-SHEQ00}).

\subsection{Parabolic bases}

\noindent
{\bf 1.2.1. Separation of variables}.

\noindent
Parabolic coordinates $\xi$ and $\eta$ are connected with the
Cartesian  $x$ and $y$ by
\begin{equation}
  x = \frac{1}{2}(\xi^2 - \eta^2)\enspace,\qquad
  y=  \xi \eta \enspace,\qquad
   \xi\in {\bf R}, \, \eta > 0.
\label{P1}
\end{equation}
The Laplacian and two-dimensional volume element are given by
\bea
\Delta =
\frac{\partial^2}{\partial x^2}
+ \frac{\partial^2}{\partial y^2}
= {1\over \xi^2 + \eta ^2}
\left(\frac{\partial^2}{\partial \xi^2}
+ \frac{\partial^2}{\partial \eta^2}\right),
\qquad
dv = dxdy = (\xi^2+\eta^2) d\xi d\eta.
\label{P01}
\eea
The Schr\"odinger equation in parabolic coordinates (\ref{P1}) is
\be
\label{V1-EQ1}
{1\over \xi^2 + \eta ^2}
\left({\partial^2 \Psi \over \partial \xi^2} +
{\partial^2 \Psi \over \partial \eta^2}\right)
+ \left[2E - \omega^2 (\xi^4 - \xi^2 \eta^2 + \eta^4) - k_1
(\xi^2 - \eta^2) - {k_2^2- {1\over 4}\over \xi ^2\eta ^2}\right]
\Psi = 0.
\ee
Upon substituting
$$
\Psi (\xi, \eta) = X (\xi) Y (\eta)
$$
and introducing the parabolic separation constant $\lambda$, the
equation (\ref{V1-EQ1}) split into two ordinary differential equations:
\bea
\label{V1-EQ2}
\frac{d^2 X}{d\xi^2}
&+& \left(2E \xi^2 - \omega^2\xi^6 - k_1 \xi^4
- \frac{k_2^2 -\frac{1}{4}}{\xi^2}\right)X
= - \lambda X,
\\[2mm]
\frac{d^2 Y}{d\eta^2}
&+& \left(2E \eta^2 - \omega^2\eta^6 + k_1 \eta^4
- \frac{k_2^2 -\frac{1}{4}}{\eta^2}\right) Y
= + \lambda Y.
\label{V1-EQ-2}
\eea
Equations (\ref{V1-EQ2}) and (\ref{V1-EQ-2}) are transformed into one
another by change $\xi  \longleftrightarrow  i\eta$. We have
\be
\label{V1-EQ002}
\Psi (\xi, \eta; E, \lambda) = C(E,\lambda)
Z (\xi; E,\lambda) Z (i\eta; E, \lambda)
\ee
where $C(E,\lambda)$ is the normalization constant determined
by the condition
\be
\label{V1-EQ102}
\int_{0}^{\infty}d\eta\int_{-\infty}^{\infty} d\xi (\xi^2+\eta^2)
|\Psi (\xi, \eta; E, \lambda)|^2  = 1
\ee
and the function $Z (\mu; E,\lambda)$ is a solution of the equation
\bea
\label{V1-EQ202}
\left[- \frac{d^2 }{d\mu^2}
+ \left(\omega^2\mu^6 + k_1 \mu^4 - 2E \mu^2
+ \frac{k_2^2 -\frac{1}{4}}{\mu^2}\right)\right] Z (\mu; E,\lambda)
= \lambda Z (\mu; E,\lambda).
\eea
Thus, at $\mu \in (-\infty, \infty)$ we have eq. (\ref{V1-EQ2}) and
at $\mu\in [0, i \infty)$ - the eq. (\ref{V1-EQ-2}). Note that
in the complex $\mu$ domain the ``physical'' region is just the two lines
${\rm Im}\  \mu=0$ and ${\rm Re}\  \mu = 0, {\rm Im}\  \mu>0$.
Our task is to find the solutions of eq. (\ref{V1-EQ202}) that are regular and decreasing as $\mu\to \pm\infty$
and $\mu\to i\infty$ .

\vspace{0.4cm}
\noindent
{\bf 1.2.2. Recurrence relations}.

\noindent
Consider now the equation (\ref{V1-EQ202}). To solve it we make
the substitution
\be
\label{V1-EQ4}
Z(\mu; E,\lambda) = \exp \left(-\frac{\omega}{4}\mu^4 -
\frac{k_1}{4\omega}\mu^2 \right)
\,
\mu^{\frac12\pm k_2}
\,
\psi (\mu; E,\lambda),
\ee
and obtain the differential equation
\be
\label{V1-EQ5}
\frac{d^2 \psi}{d\mu^2} +
\left[\frac{2(\frac12\pm k_2)}{\mu} -
2\omega\mu\left(\mu^2+\frac{k_1}{2\omega^2}\right)\right]
\frac{d \psi}{d\mu} +
\left[2{\tilde{E}}\mu^2 + {\tilde{\lambda}}\right] \psi = 0
\ee
where
\be
\label{V1-EQ6}
{\tilde{E}} =  E + \frac{k_1^2}{8\omega^2} - \omega
(2\pm k_2),
\qquad
{\tilde{\lambda}} = \lambda -  \frac{k_1}{\omega}(1 \pm k_2).
\ee
Passing to a new variable $z=\mu^2$ in eq. (\ref{V1-EQ5}),
we have
\be
\label{V1-EQ7}
z \frac{d^2 \psi}{d z^2} +
\left[(1\pm k_2) - \omega z \left(z +
\frac{k_1}{2\omega^2}\right)\right]
\frac{d \psi}{d z} +
\left[\frac12{\tilde{E}} z  +
\frac14{\tilde{\lambda}}\right] \psi = 0
\ee
We express the wave function $\psi(z)$ in the  form
\be
\label{V1-EQ8}
\psi(z; E,\lambda) = \sum_{s=0}^{\infty} A_s(E,\lambda) z^s
\ee
The substitution (\ref{V1-EQ8}) in eq. (\ref{V1-EQ7}) leads to
the following three-term recurrence relation for the
expansion coefficients $A_s \equiv A_s(E,\lambda)$,
\bea
\label{V1-EQ9}
(s+1)(s+1 \pm k_2) A_{s+1}
&+& \frac14 \biggl[\lambda - \frac{k_1}{\omega}
(2s + 1 \pm k_2) \biggr] A_{s}
\nonumber\\[2mm]
&+&\frac12\biggl[E + \frac{k_1}{8\omega^2}
- \omega (2s \pm k_2) \biggr] A_{s-1} = 0,
\eea
with the initial conditions $A_{-1} = 0$ and $A_{0}=1$.

\vspace{0.4cm}
\noindent
{\bf 1.2.3. Asymptotic behavior}.

\noindent
To understand the  behavior of the coefficients for large $s$ we use
continued fractions theory \cite{LORENTZEN}. Setting
\bea
\label{V1-EQ9a} \frac{A_s}{A_{s-1}}&=&\xi_sf(s)\\
f(s)&=& \sqrt{\frac{\omega}{2}}\frac{\Gamma(\frac{s}{2}
+\frac12)\Gamma(\pm\frac{k_2}{2}
+\frac{s}{2}+\frac12)\Gamma(\frac{\alpha}{2\omega}
\pm\frac{k_2}{2}+\frac{s}{2}+1)}
{\Gamma(\frac{s}{2}+1)\Gamma(\pm\frac{k_2}{2}
+\frac{s}{2}+1)\Gamma(\frac{\alpha}{2\omega}
\pm\frac{k_2}{2}+\frac{s}{2}+\frac12)}.\nonumber
\eea
where $\Gamma(z)$ is the  Gamma function, we can write the recurrence
relation (\ref{V1-EQ9}) in the standard form
\bea  \label{V1-EQ9b} \xi_s&=&\frac{1}{b_s+\xi_{s+1}},\\
b_s&=& \sqrt{\frac{2}{\omega}}\frac{\Gamma(\frac{s}{2}+\frac12)
\Gamma(\pm\frac{k_2}{2}
+\frac{s}{2}+\frac12)\Gamma(\frac{\alpha}{2\omega}
\pm\frac{k_2}{2}+\frac{s}{2}+1)[-\lambda+\frac{k_1}{\omega}
(2s+1\pm k_2)]}{\Gamma(\frac{s}{2}+1)\Gamma(\pm\frac{k_2}{2}
+ \frac{s}{2}+1)\Gamma(\frac{\alpha}{2\omega}
\pm\frac{k_2}{2}+\frac{s}{2}+\frac32)2^3},
\nonumber
\eea
where $\alpha=-(E+k_1/(8\omega^2))/2$. Note that
$$ f(s+1)f(s)=\omega\frac{(\frac{\alpha}{\omega}\pm
  k_2+s+1)}{(s+1)(\pm k_2+s+1)}.
$$
Stirling's formula for the  Gamma function
$\Gamma(z)=z^{z-1/2}e^{-z}\sqrt{2\pi}(1+O(1/z))$ as $|z|\to\infty$ with $|{\rm arg}\
z|<\pi$, gives $f(s)=-\sqrt{\frac{\omega}{s}}(1+O(1/s))$ and
$b_s=\pm\frac{k_2}{2\omega\sqrt{s\omega}}(1+O(1/s))$. In the following
  we take $k_1\le 0$, $k_2, \lambda, E$ real and $\omega>0$. Without loss of
  generality we  can assume $b_s$ is positive for sufficiently large
  $s$ since, otherwise, we could make the replacements $b_s\to-b_s,\
  \xi_s\to-\xi_s$.

Since $\sum b_s=\infty$, it is a consequence of the Seidel-Stern theorem
that the formal continued fraction expressions for the $\xi_s$ converge:
$$
\xi_s\qquad =
\qquad
\stackrel{1}{\overline{b_s}}
\stackrel{}{+}\stackrel{1}{\overline{b_{s+1}}}
\stackrel{}{+}\stackrel{}{\cdots}
\stackrel{}{+}\stackrel{1}{\overline{b_{s+k}}}
\stackrel{}{+}\stackrel{}{\cdots}.
$$
Moreover, standard continued fraction theory tells us that
\be
\label{V1-EQ9c} \xi_s=\lim_{n\to\infty}\frac{A^{(s)}_n}{B^{(s)}_n}
\ee
where
\be
\label{V1-EQ9d}
\left(\ba{ll} A^{(s)}_{-1}\\ B^{(s)}_{-1}\ea\right)=\left(\ba{ll} 1\\
0\ea\right),
\quad
\left(\ba{ll} A^{(s)}_0\\ B^{(s)}_0\ea\right)=\left(\ba{ll} 0\\
1\ea\right),
\ee
and
$$
\left(\ba{ll} A^{(s)}_n\\ B^{(s)}_n\ea\right)=b_{n+s}\left(\ba{ll}
A^{(s)}_{n-1}\\ B^{(s)}_{n-1}\ea\right)+\left(\ba{ll}
A^{(s)}_{n-2}\\ B^{(s)}_{n-2}\ea\right),\quad n\ge 1.
$$
Furthermore the relation $
A^{(s)}_nB^{(s)}_{n-1}-A^{(s)}_{n-1}B^{(s)}_n=(-1)^{n-1}$ holds for
all $n\ge 0$, which implies
$$
\frac{A^{(s)}_{n}}{B^{(s)}_{n}}-\frac{A^{(s)}_{n-1}}{B^{(s)}_{n-1}}
=\frac{(-1)^{n-1}}{B^{(s)}_{n-1}B^{(s)}_{n}}.
$$
This result in turn implies that the sequence
$A^{(s)}_{2n}/B^{(s)}_{2n}$ is, for large $s$ and $n$ monotone
  increasing in $n$ and goes to $\xi_s$ in the limit, whereas
  $A^{(s)}_{2n+1}/B^{(s)}_{2n+1}$ is monotone decreasing in $n$
and goes to $\xi_s$ in the limit. For example,
\be \label{V1-EQ9e}
\frac{A^{(s)}_{2n+2}}{B^{(s)}_{2n+2}}-\frac{A^{(s)}_{2n}}{B^{(s)}_{2n}}
=\frac{b_{2n+2+s}}{B^{(s)}_{2n}B^{(s)}_{2n+2}}.
\ee
It follows from (\ref{V1-EQ9c}), (\ref{V1-EQ9e}) that
\be \label{V1-EQ9f}
\xi_s=\frac{A^{(s)}_{0}}{B^{(s)}_{0}}+\sum_{n=1}^\infty
  \frac{b_{2n+2+s}}{B^{(s)}_{2n}B^{(s)}_{2n+2}}.
\ee
Simple estimates using the recurrence relations (\ref{V1-EQ9d}) give
$$ B^{(s)}_{2n}>1+b_{s+1}\sum_{m=1}^nb_{2m+s},\quad
B^{(s)}_{2n+1}>\sum_{m=0}^nb_{2m+1+s},
$$
Substituting these results into the identities
$$ B^{(s)}_{2n}=\sum_{m=1}^nb_{2m+s}B^{(s)}_{2m-1},\quad
B^{(s)}_{2n+1}=\sum_{m=0}^nb_{2m+s+1}B^{(s)}_{2m}
$$
we get refined upper bounds for  $B^{(s)}_{2n}, B^{(s)}_{2n+2}$. We
can approximate the sum $\sum_{m=s}^n 1/\sqrt{m}$ by the
integral $\int_s^n\frac{1}{\sqrt{x}}\ dx$ and use similar
approximations to get an upper bound for the series (\ref{V1-EQ9f}):
$$
|\xi_s|<\kappa_1 \int_0^\infty\frac{dy}{\sqrt{y+s}(y^2+1)}+\kappa_2
$$
for  positive constants $\kappa_j$ independent of $s$. This shows
that  $|\xi_s|$ is uniformly bounded in $s$. Since
$\xi_{s+1}=-b_s+1/\xi_s$ and $b_s\to 0$ as $s\to\infty$ it is also
true that $|1/\xi_s|$ is uniformly bounded in $s$.

It follows from (\ref{V1-EQ9b}) that
$$
\xi_{s+1}-\xi_{s-1}=\frac{\xi_{s-1}-(b_{s}+
\xi_{s-1})(1-b_{s-1}\xi_{s-1})}{1-b_{s-1}\xi_{s-1}}.
$$
Now choose $s_0$ so large that $b_{s+1}<b_s$ and $b_s\xi_s<1$ for all
$s\ge s_0$. Note from this identity that if $\xi_{s_1-1}>1$ for some
$s_1>s_0$ then $\xi_{s_1+1}>\xi_{s_1-1}>1$. Thus the sequence
$\xi_{s_1+2k-1}$ is  monotonically increasing
for all $k\ge 0$. Since $|\xi_s|$ is bounded, it follows that in this case
$\lim_{k\to\infty}\xi_{s_1+2k-1}=\xi_{+}$
exists, and $\xi_+>1$. Since $\xi_{s+1}=-b_s+1/\xi_s$, $b_s\to 0$ as
$s\to\infty$ and $|1/\xi_s|$ is uniformly bounded in $s$, then
the sequence $\xi_{s_1+2k}$ is also convergent,
$\lim_{k\to\infty}\xi_{s_1+2k}=\xi_{-}$
where  $0< \xi_- <  1$.

The other possibility is that $\xi\le 1 $ for all $s\ge s_0$. Since
$1/\xi_s-\xi_{s+1}=b_s\to 0$ as $s\to\infty$, and $1/\xi_s\ge 1$,
$\xi_{s+1}\le 1$ for all $s\ge s_0$ it follows that $\lim_{k\to
  \infty}\xi_k=\xi_+=\xi_-=1.$ Thus in all cases the sequences
$\xi_{2k}$ and $\xi_{2k+1}$ converge.

We conclude that
$$
\frac{A_{s+1}}{A_s}=f(s+1)\xi_s=
\sqrt{\frac{\omega\xi_\pm}{s}}(1+O(1/s)),\quad \xi_+\xi_-=1,
$$
depending on whether $s$ is even or odd. Thus asymptotically  $A_s\sim
\sqrt{ \xi_\pm}\sqrt{\frac{\omega^s}{s!}}$, depending on whether $s$ is
even or odd, and
\bea
\label{V1-EQ13}
\psi(z) \sim \sum \frac{\sqrt{\xi_\pm}( \sqrt{\omega}z)^{s}}{\sqrt{s!}}.
\eea
Then we have for $z>0$ [the case of  eq. (\ref{V1-EQ2})]
\bea
\label{V1-EQ14}
\sum \frac{\sqrt{\xi_\pm}(\sqrt{\omega}z)^{s}}{\sqrt{s!}} >
\sqrt{\sum \frac{\xi_\pm(\omega z^2)^{s}}{s!}} =
\xi_\pm \cosh \left(\frac{\omega}{2} z^2 \right)+\xi_
\mp  \sinh \left(\frac{\omega}{2} z^2 \right).
\eea
This function does not belong to the Hilbert space. If $k_1>0$ then we
must make the replacements $b_s\to -b_s$ and $\xi_s\to-\xi_s$. This
has the effect of replacing $z$ by $-z$ in (\ref{V1-EQ13}). Now the
asymptotic solution is oscillatory. However, then for  $z<0$ [the case
of eq. (\ref{V1-EQ-2})] the solution doesn't belong to the Hilbert space.
The solution we have found is the minimal solution of the three-term
recurrence relations. There is a linearly independent solution, but the
coefficients grow more rapidly than the minimal solution coefficients.

\vspace{0.4cm}
\noindent
{\bf 1.2.4. Energy spectrum and separation constant}.

\noindent
Thus the function $Z(\mu)$ cannot converge simultaneously at
large $\mu$ for real and imaginary $\mu$ and therefore the series
(\ref{V1-EQ8}) should be truncated. The condition for series
(\ref{V1-EQ8}) to be truncated results in the energy spectrum
giving the same formula (\ref{V2-E1}) and now the coefficients
$A_s\equiv A_s^{n q} (k_1, \pm k_2)$ satisfy the following
 relation
\bea
\label{V1-EQ17}
(s+1)(s+1 \pm k_2) A_{s+1}
+ \beta_s  A_{s}
+ \omega(n+1-s) A_{s-1} = 0,
\eea
$$
\beta_s
= \frac{\lambda}{4} - \frac{k_1}{4\omega}
(2s + 1 \pm k_2).
$$
The three-term recurrence relation (\ref{V1-EQ17}) represents a
homogeneous system of $n+1$ - algebraic equations for $n+1$ - coefficients
$\{A_0, A_1, A_2, ... A_n\}$. The requirement for the existence of a non-trivial
solution leads to a vanishing of the determinant
\bea
\label{V1-EQ18}
D_n (\lambda) =
\left|
\begin{array}{ccccccc}
\beta_0&1\pm k_2&&\cdot&& \\
\omega n&\beta_1&2(2\pm k_2)&\cdot&&\\
\cdot&\cdot&\cdot&\cdot&&\cdot&\cdot\\
&&&\cdot&2\omega&\beta_{n-1}&n(n\pm k_2)\\
&&&\cdot&&\omega&\beta_n
\end{array}
\right| = 0
\eea
The roots of the corresponding algebraic equation give us the $(n+1)$
eigenvalues of the parabolic separation constant $\lambda_n (k_1, \pm k_2)$.
It is known that all roots for a such determinants are real and
distinct \cite{COJO}. Thus  all  values of the separation constant
are real and can be enumerated with the help of the integer  $q$,
namely the values are $\lambda_n (k_1, \pm k_2) \to \lambda_{nq} (k_1, \pm k_2)$, where
$0 \leq q \leq n$. The degeneracy for the $n$ - energy
state, as in the Cartesian case, equals $n+1$.

Note that eq. (\ref{V1-EQ18}) is invariant under
the simultaneous transformation $k_1 \to - k_1$ and $\lambda \to -\lambda$.
Thus  if one of the $\lambda=\lambda_{n} (k_1, \pm k_2)$ is
root of eq. (\ref{V1-EQ18}), then $\lambda= -\lambda_{n} (-k_1, \pm k_2)$
 is also a root of the same equation. We see that for the odd
energy state ($n$-odd) the range of $\lambda_{nq} (k_1, \pm k_2)$ splits
into two subset $\lambda^{(1)}_{nq}$ and $\lambda^{(2)}_{nq}$ connected
with each to other by the relation
$\lambda_{nq}^{(1)} (k_1, \pm k_2) \longleftrightarrow
- \lambda_{nq'}^{(2)} (-k_1, \pm k_2)$.
For $n$-even, always there exists the additional root
$\lambda_{nq}(k_1, \pm k_2) = - \lambda_{nq}(-k_1, \pm k_2)$, which
equals zero when $k_1=0$.

\vspace{0.4cm}
\noindent
{\bf 1.2.5. Wave functions}.

\noindent
We will term the polynomial solutions of eq.(\ref{V1-EQ7}), or
eq.(\ref{V1-EQ5}), as $Mk_{nq} (z; k_1, \pm k_2)$, and the
function (\ref{V1-EQ4}) as $Ta_{nq} (z; k_1, \pm k_2)$
\footnote {The notation $Ta$ - devoted to the memory of Professor
V.Ter-Antonyan (1942-2003)}.
Then the physical admissible solutions of eq. (\ref{V1-EQ7}) have
the form
\bea
\label{V1-EQ0-8}
Mk_{nq} (z; k_1, \pm k_2) \equiv \psi (z; E, \lambda) =
\,
\sum_{s=0}^{n} A_s^{n q} (k_1, \pm k_2) \, z^{s},
\eea
and the corresponding solution of eq. (\ref{V1-EQ4}) is
\bea
\label{V1-EQ20}
Ta_{nq} (\mu; k_1, \pm k_2) =
exp \left(-\frac{\omega}{4}\mu^4 -
\frac{k_1}{4\omega}\mu^2 \right)
\,
\mu^{\frac12\pm k_2}
\,
Mk_{nq} (\mu^2; k_1, \pm k_2)
\eea
Observe that parabolic wave functions (as also Cartesian wave functions)
split into two classes and transform to each other via  $k_2 \to - k_2$.
In the case  $k_2 =0$ (when the centrifugal term disappears),  the solution
(\ref{V1-EQ20}) becomes an even and odd parity wave function under
exchange $\mu \to - \mu$.

It is known that there exists a direct connection between the
quantum numbers $q$ and numbers of zeros of the polynomial (\ref{V1-EQ0-8})
and, therefore, the eigenvalues of the separation constant
$\lambda_{nq}(k_1, \pm k_2)$ may be ordered by the numbers of nodes of
the wave function $Ta_{nq} (\mu; k_1, \pm k_2)$.
Indeed by we will see that these are orthogonal polynomials, hence
\cite{INCE},  all the $n$ - zeros of the $Mk_{nq} (z; k_1, \pm k_2)$ are situated on the real axis
$-\infty < z < \infty$, and all zeros have multiplicity one. Assume that the separation constants
$\lambda_{nq}(k_1, \pm k_2)$ are numerated in ascending order,
i.e
\bea
\label{V1-EQ21}
\lambda_{n0}(k_1, \pm k_2) < \lambda_{n1}(k_1, \pm k_2) <
......... < \lambda_{n,n-1}(k_1, \pm k_2) < \lambda_{n,n}(k_1, \pm k_2)
\eea
then according to the oscillation theorem \cite{HILBERT}, the quantum number
$q$ also enumerates the zeros of polynomials $Mk_{nq} (z; k_1, \pm k_2)$ in
the region $z > 0$, or the real axis of $\mu$, (see eqs. (\ref{V1-EQ0-80})
- (\ref{V1-EQ0-87})). Let us now  introduce two quantum number $q_1$ and
$q_2$, which determine the zeros of polynomials $Mk_{nq} (z; k_1, \pm k_2)$
for $z > 0$ and $z < 0$, correspondingly. Then $q_1+q_2=n$, and
\bea
\label{V1-EQ22}
\lambda_{nq_1}(k_1, \pm k_2) = - \lambda_{nq_2}(-k_1, \pm k_2)
\eea
For $\mu = \xi$ the function (\ref{V1-EQ20})  gives the solution of
equation (\ref{V1-EQ2}), and for  $\mu = i\eta$ the solution for
equation (\ref{V1-EQ-2}).
Thus the parabolic wave function (\ref{V1-EQ002}) can be written in
following way
\bea
\label{V1-EQ23}
\Psi_{n q_1 q_2} (\xi, \eta; k_1, \pm k_2) =
C_{n q_1 q_2}(k_1, \pm k_2) \,
Ta_{nq_1} (\xi; k_1, \pm k_2) \,
Ta_{nq_2} (i\eta; k_1, \pm k_2).
\eea

\vspace{0.4cm}
\noindent
{\bf 1.2.7. The particular cases}.

\noindent
Let us consider some low energy state $n=0,1,2$.  In case of $n=0$
we have
\bea
\label{V1-EQ0-80}
Mk_{00} (z) = 1,
\qquad
\lambda_{00} = \frac{k_1}{\omega} (1\pm k_2)
\eea
For $n=1$ we have that $q=0,1$ and equation (\ref{V1-EQ18}) admits two
solutions
\bea
\label{V1-EQ0-81}
\lambda_{10} = \frac{k_1}{\omega} (2\pm k_2) -
\sqrt{\frac{k_1^2}{\omega^2} + 16\omega (1\pm k_2)},
\quad
\lambda_{11} = \frac{k_1}{\omega} (2\pm k_2) +
\sqrt{\frac{k_1^2}{\omega^2} + 16\omega (1\pm k_2)}.
\eea
Therefore there are two polynomials
\bea
\label{V1-EQ0-82}
Mk_{10} (z; k_1, \pm k_2)
&=&
1 - \frac{k_1- \sqrt{k_1^2 + 16\omega^3(1\pm k_2)}}
{4\omega(1\pm k_2)} z,
\\
Mk_{11} (z; k_1, \pm k_2)
&=&
1 - \frac{k_1 + \sqrt{k_1^2 + 16\omega^3(1\pm k_2)}}
{4\omega(1\pm k_2)} z.
\eea
For $n=2$ the equation (\ref{V1-EQ18}) is equivalent to a cubic
algebraic equation
\bea
\label{V1-EQ0-83}
\lambda^3 - \frac{3k_1}{2\omega}\lambda^2 +
[\frac{k_1^2}{2\omega^2} - 2\omega (3\pm k_2)] \lambda
+ 2k_1 (1\pm k_2) = 0.
\eea
At $k_1=0$ this equation can be simplified, and we have
following three solutions
\bea
\label{V1-EQ0-84}
\lambda_{20} =  - \sqrt{32\omega (3\pm k_2)},
\qquad
\lambda_{21} = 0,
\qquad
\lambda_{22} =  \sqrt{32\omega (3\pm k_2)}
\eea
with
\bea
\label{V1-EQ0-85}
Mk_{20} (z)
&=&
1 + \frac{\sqrt{2\omega(3\pm k_2)}}{(1\pm k_2)} z +
\frac{\omega}{(1\pm k_2)} z^2
\\
\label{V1-EQ0-86}
Mk_{21} (z)
&=&
1 - \frac{\omega}{(2\pm k_2)} z^2
\\
\label{V1-EQ0-87}
Mk_{22} (z)
&=&
1 - \frac{\sqrt{2\omega(3\pm k_2)}}{(1\pm k_2)} z +
\frac{\omega}{(1\pm k_2)} z^2.
\eea

\vspace{0.4cm}
\noindent
{\bf 1.2.6. Orthogonality relations and normalization constant}.

\noindent
The wave functions (\ref{V1-EQ23}) as  eigenfunctions of Hamiltonians
are orthogonal for quantum number $n$, or for $n\not= n'$
\bea
\label{V1-EQ1-01}
\int_{0}^{\infty}d\eta\int_{-\infty}^{\infty} d\xi (\xi^2+\eta^2)
\Psi_{n' q_1 q_2}^{*}(\xi, \eta; k_1, \pm k_2)
\Psi_{n q_1 q_2}^{*}(\xi, \eta; k_1, \pm k_2) =  0.
\eea
Because the energy spectrum is degenerate there exist additional
orthogonality relations for quantum number $q$. Using the equations
(\ref{V1-EQ2}) and (\ref{V1-EQ-2}) it is easy to prove that for
$q_1 \not= q_1'$ and $q_2 \not= q_2'$
\bea
\label{V1-EQ1-02}
\int_{-\infty}^{\infty} d\xi
Ta_{nq_1'}^{*} (\xi; k_1, \pm k_2) \,
Ta_{nq_1} (\xi; k_1, \pm k_2)  = 0
\\[2mm]
\int_{0}^{\infty} d\eta
Ta_{nq_2'}^{*} (i\eta; k_1, \pm k_2)
Ta_{nq_2} (i\eta; k_1, \pm k_2) = 0.
\eea
Thus we have for $q \not= q'$
\bea
\label{V1-EQ1-03}
\int_{0}^{\infty}d\eta\int_{-\infty}^{\infty} d\xi (\xi^2+\eta^2)
\Psi_{n q_1' q_2'}^{*}(\xi, \eta; k_1, \pm k_2)
\Psi_{n q_1 q_2}(\xi, \eta; k_1, \pm k_2) =  0.
\eea
Let us now  calculate the normalization constant
$C_{n q_1 q_2}(k_1, \pm k_2)$. From the explicit form of the wave
function $\Psi_{n q_1 q_2}^{*}(\xi, \eta; k_1, \pm k_2)$ and the
normalization condition (\ref{V1-EQ102}), it follows that
\bea
\label{V1-EQ1-04}
\frac{1}{8}\, |C_{n q_1 q_2}(k_1, \pm k_2)|^2
\sum_{s,s',t,t'=0}^{n} (-1)^{t+t'} \,
A_s^{n} (k_1, \pm k_2)
A_{s'}^{n} (k_1, \pm k_2)
\nonumber\\[2mm]
\times
A_t^{n} (k_1, \pm k_2)A_{t'}^{n} (k_1, \pm k_2)
\,
\{F^{-1/4}_{t,t'} F^{+1/4}_{s,s'} + F^{+1/4}_{t,t'} F^{-1/4}_{s,s'}
\} = 1
\eea
where
\bea
\label{V1-EQ1-05}
F^{\pm 1/4}_{t,t'} = \sum_{m=0}^{\infty}
\frac{\Gamma(\frac{m+t+t'\pm k_2+1}{2} + \frac{1}{4} \pm \frac{1}{4})}
{m!} \left(\frac{k_1}{2\omega}\right)^m.
\eea

\subsection{Niven approach}

Let us express solution of the Schr\"odinger equation (\ref{V1-EQ0})
in the following form \cite{KMP-96}
\be
\Psi(x,y) = e^{-\omega (x + \frac{k_1}{4\omega^2})^2
- \frac12 \omega y^2} \,
y^{\frac12 \pm k_2} \, \Phi (x,y).
\ee
From the eqs. (\ref{V1-SHEQ2}), (\ref{V1-SHEQ3}) and (\ref{V1-EQ4})
follows that the function $\Phi (x,y)$ is polynomial (product of two
polynomials) from variables $(x, y^2)$ in Cartesian coordiantes and
$(\xi^2, \eta^2)$ for parabolic ones.  It satisfy the equation
\be
\label{mul02}
{\cal R} \Phi (x,y) =
- 2E \Phi (x,y),
\ee
where the operator ${\cal R}$ is
\be
\label{mul2}
{\cal R}
=
{\partial^2\over\partial x^2}
+ {\partial^2\over \partial y^2} +
\left[{(1\pm 2k_2)\over y} - 2\omega y\right]
{\partial \over \partial y} -
4\omega \left[x + \frac{k_1}{4\omega^2}\right]
{\partial \over \partial x}
- \omega (2 \pm k_2) + \frac{k_1^2}{8\omega^2}.
\ee
Taking into account that
\bea
\label{V1-EQ3-01}
Mk_{nq} (z; k_1, \pm k_2)  =
\sum_{s=0}^{n} A_s^{n q} (k_1, \pm k_2) \, z^{s}
= \Pi_{\ell=1}^{n} (z-\alpha_{\ell}),
\eea
where $\alpha_{\ell}$, $\ell = 1,2,...n$ are zeros of polynoms
$Mk_{nq}(z)$ on the real axis $-\infty < z < \infty$, and that in
parabolic coordinates
\bea
\label{V1-EQ3-02}
{y^2\over \alpha} + 2x - \alpha
= \frac{(\xi^2 - \alpha)(\eta^2+ \alpha)}{\alpha},
\eea
we can choose a solution of eq. (\ref{mul02}) in the form
\be
\label{V1-EQ3-03}
\Phi (x,y) =
Mk_{nq_1} (\xi^2; k_1, \pm k_2) \,
Mk_{nq_2} (-\eta^2; k_1, \pm k_2)
\cong
\Pi^n_{\ell = 1}\left({y^2\over \alpha_\ell } +
2x - \alpha_\ell \right).
\ee
Then from (\ref{mul02}) follows that zeros $\alpha_\ell$ must satisfy
the systems of $n$ - algebraic equations
\bea
\label{V1-EQ3-04}
\sum_{m \neq \ell}^{n} {2\over \alpha_\ell  -  \alpha_m}
+ {(1\pm k_2)\over \alpha_\ell }
- \omega \alpha_\ell = \frac{k_1}{2\omega},
\qquad
\ell = 1,2,....n,
\eea
and for energy spectrum we again have a formala (\ref{V2-E1}).
The system algebraic equation (\ref{V1-EQ3-04}) contains $n$ - set
of solutions (zeros) $(\alpha_1^{(q)}, \alpha_2^{(q)}, .....
\alpha_n^{(q)})$,  $q=1,2,...n$ and all zeros are real.
The positive zeros $\alpha_{\ell} > 0$
define the nodes of wave functions for equation (\ref{V1-EQ2}),
whereas negative zeros $\alpha_{\ell} < 0$ define the nodes of
wave functions for equation (\ref{V1-EQ-2}).

The eigenvalues of parabolic separation constant can be calculated
the same way via the operator equation
${\Lambda} \Phi (x,y) = \lambda \Phi (x,y)$ (see for details
\cite{KMP-96}). More elegant way is direct to use the differential
equation (\ref{V1-EQ7}) \cite{USH}.
Rewriting first the eq. (\ref{V1-EQ7}) in following form
\bea
\label{V1-EQ7-1}
\left\{ 4z \frac{d^2}{d z^2} +
4 \left[(1\pm k_2) - \omega z \left(z +
\frac{k_1}{2\omega^2}\right)\right]
\frac{d}{d z}
+
\left[ 4n \omega z  - \frac{k_1}{\omega}(1\pm k_2)
\right ] \right\} Mk_{nq} (z; k_1, \pm k_2)
\nonumber
\\[2mm]
=
\lambda Mk_{nq} (z; k_1, \pm k_2)
\eea
Putting now the wave function $Mk_{nq} (z; k_1, \pm k_2)$ in form
of (\ref{V1-EQ3-01}), we arrive to the following result
\bea
\label{V1-EQ3-05}
\lambda_{nq}(k_1, \pm k_2) = 4(1 \pm k_2) \left[
\frac{k_1}{4 \omega} +  \sum_{\ell=1}^{n}\frac{1}{\alpha_\ell^{(q)}}
\right],
\eea
(in case of $n=0$ the sum must be eliminated) where the quantum number
$q=1,2,..n$ labeled the eigenvalue of parabolic separation constant.

\subsection{Interbasis expansions between Cartesian
and parabolic bases}

We determine the interbasis expansion relating Cartesian and parabolic
wave functions at the fixed energy $E_n$
\bea
\label{V1-INTER1}
\Psi_{n q_1 q_2}(\xi, \eta; \pm k_2) =
\sum_{n_1=0}^{n_1+n_2}
\,
W_{n q_1 q_2}^{n_1 n_2} (\pm k_2) \,
\Psi_{n_1 n_2}(x, y; \pm k_2).
\eea
For simplicity we consider only the case when $k_1=0$. To calculate
the interbasis coefficients $W_{n q}^{n_1 n_2}$ we  use
asymptotic methods \cite{MARDO2}. First, we change in $\Psi_{n q_1 q_2}(\xi, \eta; \pm k_2)$
from parabolic coordinates to Cartesian ones
\bea
\label{V1-INTER2}
\xi^2 = \sqrt{x^2 + y^2} + x,
\qquad
\eta^2 = \sqrt{x^2 + y^2} - x,
\eea
and then let $y$ tend to zero on the both sides of (\ref{V1-INTER1}).
As a result
\bea
\label{V1-INTER3}
\xi^2 \to |x| + x,
\qquad
\eta^2 \to |x| - x,
\eea
and the dependence on variable $y$ in (\ref{V1-INTER1}) is removed.
Using  the orthogonality conditions for Hermite polynomials we
get the equality
\bea
\label{V1-INTER4}
W_{n q_1 q_2}^{n_1 n_2} (\pm k_2) =
C_{n q_1 q_2}(\pm k_2)\,
\left(\frac{2\omega}{\pi}\right)^{\frac{1}{4}}
\,
\sqrt{\frac{\Gamma(n_2 +1 \pm k_2)}{2^{n_1+1} (n_1)! (n_2)!
\omega^{(1\pm k_2)}}}
\,\,
L_{n q_1 q_2}^{n_1} (\pm k_2)
\eea
where
\bea
\label{V1-INTER5}
L_{n q_1 q_2}^{n_1} (\pm k_2)
&=&
\int_{-\infty}^{\infty} \,
Mk_{nq_1} (|x|+x; \pm k_2) Mk_{nq_1} (i|x|-ix; \pm k_2) \,
e^{-2\omega x^2} \, H_{n_1} (\sqrt{2\omega} x)\, dx
\nonumber\\[2mm]
&=&
\int_{0}^{\infty} \,
\left[ Mk_{nq_1} (2x; \pm k_2) + (-1)^{n_1}
Mk_{nq_2} (2ix; \pm k_2)\right] \,
e^{-2\omega x^2} \, H_{n_1} (\sqrt{2\omega} x)\, dx.
\nonumber
\eea
With the formula (\ref{V1-EQ0-8}) the latter integral can be
expressed via the coefficients $A_{s}^{n} (\pm k_2)$:
\bea
\label{V1-INTER6}
L_{n q_1 q_2}^{n_1} (\pm k_2)
=
\frac{1}{\sqrt{2\omega}}\,
\sum_{s=0}^{n} \left[1+ (-1)^{s+n_1}\right]\,
A_{s}^{n} (\pm k_2) \,
\left(\frac{2}{\omega}\right)^s
\,
\int_{0}^{\infty} \, z^{2s}\,
e^{- z^2} \, H_{n_1} (z)\, dz.
\eea
It is more convenient to calculate $L_{n q_1 q_2}^{n_1} (\pm k_2)$
independently for even and odd $n_1$. By integration by parts we
find that integral in the sum is nonzero only at
$\frac{n_1}{2} \leq s \leq n_1+n_2$, for $n_1$ - even and
$\frac{n_1+1}{2} \leq s \leq n_1+n_2$, for $n_1$ - odd.
By changing the sum index and making the integration we finally get
\bea
\label{V1-INTER7}
W_{n q_1 q_2}^{n_1 n_2 (\pm)} (\pm k_2)
&=&
C_{n q_1 q_2}(\pm k_2)\,
\left(\frac{\pi}{2\omega}\right)^{\frac14}\,
\sqrt{\frac{\Gamma(n_2 +1 \pm k_2)}{2\omega^{(1+n_1\pm k_2)}
(n_1)! (n_2)!}}
\nonumber\\[3mm]
&\times&
\left\{\matrix{
\sum_{s=0}^{n-\frac{n_1}{2}}
\left[1+ (-1)^{s+\frac{n_1}{2}}\right]\,
A_{s+\frac{n_1}{2}}^n\, \frac{\Gamma(2s+n_1+1)}{(2\omega)^s s!},
\cr
\cr
\sum_{s=0}^{n-\frac{n_1+1}{2}}
\left[1+ (-1)^{s+\frac{n_1-1}{2}}\right]\,
A_{s+\frac{n_1+1}{2}}^n\,
\frac{\Gamma(2s+n_1+2)}{(2\omega)^s \Gamma(s+3/2)}
\cr}\right.
\eea
for $n$ even and odd respectively.

\section{Singular circular oscillator}

The second potential of the singular oscillator is ($k_1,k_2 > 0$)
\be
V_2(x,y) = \frac{1}{2}\omega^2(x^2+y^2)+
\frac{1}{2}\left(\frac{k^2_1 - \frac{1}{4}}{x^2} +
\frac{k^2_2 - \frac{1}{4}}{y^2}\right).
\label{V2}
\ee
The corresponding Schr\"odinger equation separates in three different
orthogonal coordinate systems: {\bf Cartesian, polar and elliptical
coordinates}.

\subsection{Cartesian bases}

Let us first consider the separation of variables in Cartesian
coordinates. From the asymptotic ansatz
\bea
\Psi(x,y) =
x^{\frac12\pm k_1} \ y^{\frac12\pm k_2}
\,
{\exp}[- \omega (x^2+y^2)]
\
X(x) X(y)
\label{SCH-V2-EL00}
\eea
we obtain two independent and identical separation equations
\be
\label{SCH-V2-EL01}
\left[{\partial ^2\over \partial z_{i}^2}+
\left(-2\omega  + {{1 \pm 2k_i}\over x_{i}^2}\right)
x {\partial \over \partial x_{i}} -
(1 \pm 2k_i)\omega \right] X(x_{i}) = 2\lambda_i X(x_{i}),
\qquad i =1,2
\ee
where $x_{1}=x, x_{2}=y$ and $\lambda_{1} + \lambda_{2} = - E$.
As in the case of singular anisotropic oscillator we assume that the
positive sign at the $k_i$ has to be taken if $k_i>{1\over 2}$ and
both the positive and the negative sign must be taken if
$0<k_i<{1\over 2}$.

The last equation is exactly the equation for confluent hypergeometric
functions. The quantization rule gives
\be
\label{SCH-V2-EL02}
\lambda_i = - \omega (2n_i \pm k_i + 1),
\qquad
n_i= 0,1,2,....
\ee
and the solution of eq. (\ref{SCH-V2-EL01}) in terms of Laguerre
polynomials $X (x_i) = L^{\pm k_i}_{n_i}(\omega x_i^2)$. Thus the
corresponding set of orthonormal eigenfunctions which are normalized
in quadrant $x>0, y>0$ (on 1/4) is
\bea
\label{SCH-V2-EL03}
\Psi_{n_1, n_2}^{(\pm k_1, \pm k_2)} (x,y)
= C_{n_1, n_2}^{(\pm k_1, \pm k_2)} \,
(x)^{{1\over 2} \pm k_1}(y)^{{1\over 2} \pm k_2}
e^{- {\omega \over 2}(x^2+y^2)}L^{\pm k_1}_{n_1}(\omega x^2)
L^{\pm k_2}_{n_2}(\omega y^2)
\eea
where
\bea
\label{SCH-V2-EL04}
C_{n_1, n_2}^{(\pm k_1, \pm k_2)} =
\sqrt{\frac{\omega^{2 \pm k_1 \pm k_2}\, n_1! n_2!}
{\Gamma(n_1 \pm k_1+1)\Gamma (n_2 \pm k_2+1)}}
\eea
From (\ref{SCH-V2-EL02}) we have
\bea
\label{EL-EN1}
E_n = \omega(2n + 2 \pm k_1 \pm k_2),
\eea
where $n = n_1 + n_2 = 0,1,2,...$ is principal quantum number and
the degree of degeneracy is $n+1$.

\subsection{Polar bases}

The separation of variables in the Schr\"odinger equation for
potential (\ref{SCH-V2-EL00}) in polar coordinates
\bea
\label{SCH-V2-EL-05}
x = r \cos \phi,  \qquad
y = r \sin\phi,   \qquad
0 \leq r < \infty, \,\,\,\,
0 \leq \phi < 2\pi
\eea
gives us the orthonormal solution in polynomial form
\bea
\label{SCH-V2-EL05}
\Psi^{(\pm k_1, \pm k_2)}_{n_r, m}(r, \phi)
&=&
\sqrt{{2 \omega {n_r}!\over \Gamma(n_r+2m \pm k_1 \pm k_2 + 2)}}
\,\,
(\sqrt{\omega}r)^{(2m \pm k_1 \pm k_2+1)}
\nonumber\\[2mm]
&\times& e^{-\frac{\omega r^2}{2}} \,
L^{2q \pm k_1 \pm k_2+1}_{n_r}(\omega r^2) \,
\Phi^{(\pm k_1, \pm k_2)}_m (\phi),
\qquad
n_r, m = 0,1,2,...
\\[3mm]
\label{SCH-V2-EL06}
\Phi^{(\pm k_1, \pm k_2)}_m(\phi)
&=& \sqrt{{(2m \pm k_1 \pm k_2+1)
q!\Gamma(q \pm k_1 \pm k_2 + 1)\over 2\Gamma(m \pm k_2+1)
\Gamma(m \pm k_1+1)}}
\nonumber\\[2mm]
&\times& (\cos\phi)^{1/2 \pm k_1}(\sin\phi)^{1/2 \pm k_2}
P_m^{(\pm k_2, \pm k_1)}(\cos2\phi)
\eea
where $P_m^{(\alpha,\beta)}(x)$ is a Jacobi polynomials and
$E = \omega(2n \pm k_1 \pm k_2 + 2)$, with $n = n_r + m$ and with the
same degree of degeneracy $(n+1)$.

Thus we have seen that quantum system (\ref{V2}) is {\bf exactly-solvable}
in the Cartesian and polar systems of coordinates.

\vspace{1.2cm}
\unitlength=1.00mm
\special{em:linewidth 0.4pt}
\linethickness{0.4pt}
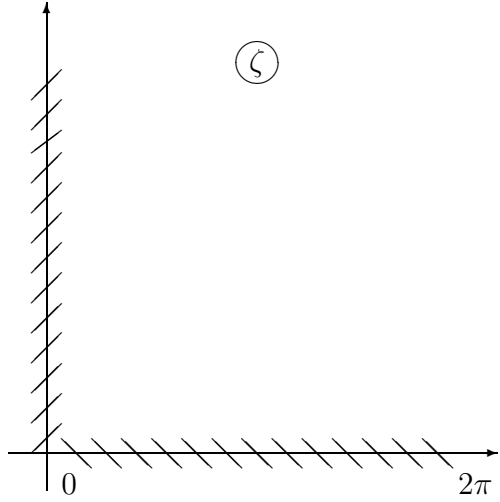
\begin{picture}(105.00,120.00)
\put(40.00,60.00){\vector(1,0){65.00}}
\put(45.00,55.00){\vector(0,1){65.00}}
\put(48.00,56.00){\makebox(0,0)[cc]{0}}
\put(47.00,62.00){\line(1,-1){4.00}}
\put(51.00,62.00){\line(1,-1){4.00}}
\put(55.00,62.00){\line(1,-1){4.00}}
\put(59.00,62.00){\line(1,-1){4.00}}
\put(63.00,62.00){\line(1,-1){4.00}}
\put(67.00,62.00){\line(1,-1){4.00}}
\put(71.00,62.00){\line(1,-1){4.00}}
\put(75.00,62.00){\line(1,-1){4.00}}
\put(79.00,62.00){\line(1,-1){4.00}}
\put(83.00,62.00){\line(1,-1){4.00}}
\put(87.00,62.00){\line(1,-1){4.00}}
\put(91.00,62.00){\line(1,-1){4.00}}
\put(95.00,62.00){\line(1,-1){4.00}}
\put(102.00,56.00){\makebox(0,0)[cc]{$2\pi$}}
\put(43.00,60.00){\line(1,1){4.00}}
\put(43.00,64.00){\line(1,1){4.00}}
\put(43.00,68.00){\line(1,1){4.00}}
\put(43.00,72.00){\line(1,1){4.00}}
\put(43.00,76.00){\line(1,1){4.00}}
\put(43.00,80.00){\line(1,1){4.00}}
\put(43.00,84.00){\line(1,1){4.00}}
\put(43.00,88.00){\line(1,1){4.00}}
\put(43.00,92.00){\line(1,1){4.00}}
\put(43.00,96.00){\line(1,1){4.00}}
\put(43.00,100.00){\line(4,3){4.00}}
\put(43.00,103.00){\line(1,1){4.00}}
\put(43.00,107.00){\line(1,1){4.00}}
\put(73.00,112.00){\circle{6.00}}
\put(73.00,112.00){\makebox(0,0)[cc]{$\zeta$}}
\end{picture}

\vspace{-4.5cm}

\subsection{Elliptic bases}

\vspace{0.2cm}
\noindent
{\bf 3.3.1. Separation of variables}.

\noindent
The elliptic coordinate $(\nu, \mu)$ connected with Cartesian one
by  ($ 0\leq \nu < \infty, \, 0\leq \mu < 2\pi $)
\begin{eqnarray}
\label{SCH-V2-EL70}
x = \frac{D}{2} \cosh\nu \cos\mu,
\qquad
y = \frac{D}{2} \sinh\nu \sin\mu,
\end{eqnarray}
where $D$ is the interfocal distance. As $D\to 0$ and $D\to \infty$,
the elliptic coordinate degenarate into the polar and Cartesian
coordinates
\bea
\label{SCH-V2-EL71}
\cosh\nu \to \frac{2r}{D},
\qquad
\cos\mu \to \cos\phi,
\qquad\qquad
(D\to 0),
\\[2mm]
\label{SCH-V2-EL710}
\sinh\nu \to \frac{2y}{D},
\qquad
\cos\mu \to \frac{2x}{D},
\qquad\qquad
(D\to \infty).
\eea
The Laplacian and volume element are
\bea
\label{SCH-V2-EL72}
\Delta =
\frac{8}{D^2(\cosh 2\nu - \cos 2\mu)}
\left(\frac{\partial^2}{\partial \nu^2}
+ \frac{\partial^2}{\partial \mu^2}\right),
\qquad
dV = \frac{D^2}{8} (\cosh 2\nu - \cos 2\mu)
d\nu d\mu.
\eea
The Schr\"odinger equation with (\ref{V2}) can be rewritten as
\bea
\label{SCH-V2-EL07}
\frac{\partial^2 \psi}{\partial \nu^2} +
\frac{\partial^2 \psi}{\partial \mu^2}
&+&
\biggl\{ \frac{D^2 E}{4} (\cosh 2\nu - \cos 2\mu) -
\frac{D^4 \omega^2}{64} (\cosh^2 2\nu - \cos^2 2\mu)
\nonumber
\\[2mm]
&-&
\left[\frac{(k^2_1 - \frac{1}{4})}{\cos^2\mu} +
\frac{(k^2_2 - \frac{1}{4})}{\sin^2\mu}\right] -
\left[\frac{(k^2_1 - \frac{1}{4})}{\sinh^2\nu} -
\frac{(k^2_2 - \frac{1}{4})}{\cosh^2\nu}\right]
\biggr\}\psi = 0.
\eea
and after the separation ansatz
\bea
\psi (\nu, \mu; D^2) = X(\nu; D^2) Y(\mu; D^2)
\label{SCH-V1-EL2}
\eea
transforms to two ordinary differential equations
\bea
\label{SCH-V1-EL3}
\frac{d^2 X}{d\nu^2}
&+&
\left[ \frac{D^2 E}{4}
\cosh 2\nu - \frac{D^4 \omega^2}{64} \cosh^2 2\nu
- \frac{k^2_2 - \frac{1}{4}}{\sinh^2\nu} +
\frac{k^2_1 - \frac{1}{4}}{\cosh^2\nu}
\right]X = - \lambda (D^2) X,
\\[3mm]
\label{SCH-V1-EL4}
\frac{d^2 Y}{d\mu^2}
&-&
\left[\frac{D^2 E}{4} \cos 2\mu - \frac{D^4 \omega^2}{64}\cos^2 2\mu +
\frac{k^2_1 - \frac{1}{4}}{\cos^2\mu} +
\frac{k^2_2 - \frac{1}{4}}{\sin^2\mu}
\right] Y = + \lambda (D^2) Y,
\eea
where $\lambda$ is the elliptic separation constant. These
equations can be written in the unit form
\bea
\label{SCH-V1-EL5}
\frac{d^2 Z(\zeta)}{d\zeta^2}
+
\left[\frac{D^4 \omega^2}{64}\cos^2 2\zeta -
\frac{D^2 E}{4} \cos 2\zeta -
\frac{k^2_1 - \frac{1}{4}}{\cos^2\zeta} -
\frac{k^2_2 - \frac{1}{4}}{\sin^2\zeta} \right] Z(\zeta) =
\lambda(D^2) Z(\zeta)
\eea
where at $\zeta\in[0,2\pi]$ we have the equation (\ref{SCH-V1-EL4})
but at $\zeta\in[0, i\infty)$ - equation (\ref{SCH-V1-EL3}). In other
words, as we see from Fig. 1 , in the complex plane $\zeta$ the "physical"
are only the shaded domain on the two lines Im $\zeta$ = 0 and
Re $\zeta$ = 0.

For $k_{1,2} > \frac12$ the centrifugal barrier is repulsive and motion
takes place in only one of the quadrants, as $\zeta\in[0,\pi/2]$, whereas for
$0 < k_{1,2} < \frac12$ it takes place in the whole region $\zeta\in[0,2\pi]$.
For the particular case $k_1=k_2= \frac12$ the equation (\ref{SCH-V1-EL5})
transforms to the problem of the ordinary two dimensional oscillator and has
been investigated in detail in the paper \cite{MPSTAN}.
In this article have shown that the solution of eq. (\ref{SCH-V1-EL5})
(for $k_1=k_2=1/2$) is described by the Ince polynomials \cite{ARS}.

In the case  where $k_1$ and $k_2$ are integers,
eqs. (\ref{SCH-V1-EL3}) and (\ref{SCH-V1-EL4}) coincide with those that
have been found via separation of
variables in the Schr\"odinger equation for the four dimensional isotropic
oscillator in spheroidal coordinates \cite{DAVTYAN1}.

\vspace{0.4cm}
\noindent
{\bf 3.3.2. Recurrence relations}.

\noindent
Let us now consider the equation (\ref{SCH-V1-EL5}).
First, introducing the function $W(\zeta; D^2)$ according to
\bea
Z (\zeta; D^2) =
{\exp}\left[- \frac{D^2\omega}{16} \cos 2\zeta\right]
W(\zeta; D^2),
\label{SCH-V1-EL22}
\eea
we have the equation
\bea
\label{SCH-V2-EL5}
\frac{d^2 W}{d\zeta^2}
+
\frac{D^2\omega}{4}\sin2\zeta
\frac{dW}{d\zeta}
+
\left[\frac{D^2\omega}{4}\cos 2\zeta -
\frac{D^2 E}{2}\cos^2\zeta -
\frac{k^2_1 - \frac{1}{4}}{\cos^2\zeta} -
\frac{k^2_2 - \frac{1}{4}}{\sin^2\zeta} - \lambda
\right] W = 0.
\eea
For $k_1=k_2=1/2$ this is the Ince equation, \cite{INCE}.

Next the substitution
\bea
W (\zeta; D^2) =
(\sin\zeta)^{\frac12\pm k_2} \ (\cos\zeta)^{\frac12\pm k_1}
\, U (\zeta; D^2)
\label{SCH-V1-EL23}
\eea
yields the equation
\bea
\frac{d^2 U}{d\zeta^2}
+
[(1\pm 2k_2) \cot\zeta - (1\pm 2k_1) \tan\zeta +
\frac{D^2\omega}{4}\sin2\zeta]
\frac{dU}{d\zeta}
+
[p \cos^2\zeta - {\tilde{\lambda}}] U = 0,
\label{SCH-V2-EL50}
\eea
where
\bea
p = \frac{D^2}{2}[\omega (2\pm k_1 \pm k_2) - E],
\quad
{\tilde{\lambda}} =
\lambda + \frac{D^2\omega}{2}(1 \pm k_1) + (1 \pm k_1 \pm k_2)^2
- \frac{D^2 E}{4} - \frac{D^4 \omega^2}{64}.
\label{SCH-V2-EL500}
\eea
Passing to a new variable $t=\cos^2\zeta$ we find
\bea
t(1-t)\frac{d^2 U}{dt^2}
+
\left\{(1\pm k_1)(1 - t) - (1\pm k_2)t
+ \frac{D^2\omega}{4} t(t-1) \right\}
\frac{dU}{dt}
+
\frac{1}{4}[p t - {\tilde{\lambda}}] U = 0,
\label{SCH-V2-EL501}
\eea
Finally, looking for the solution of the last equation in the form
\bea
\label{SCH-V1-EL24}
U (t; D^2) =
\,
\sum_{s=0}^{\infty} A_s (D^2) t^s,
\eea
for coefficients $A_{s}(D^2)$ we have the three-term recurrence
relation
\bea
\label{SCH-V1-EL204}
(s+1)(s+1 \pm k_1) A_{s+1}
&-&
\left[s(s+1\pm k_1\pm k_2) +
\frac{D^2 \omega}{4} s + \frac{{\tilde{\lambda}}}{4}\right]
A_{s}
\nonumber
\\[2mm]
&+&
\frac14 [p+ {D^2\omega}(s-1)] A_{s-1} = 0,
\eea
with $A_1 = 0$ and initial condition $A_0 = 1$.

\vspace{0.4cm}
\noindent
{\bf 3.3.3. Energy spectrum and separation constant}.

\noindent
In analogy with our asymptotic solution of the recurrence relation
for the singular anisotropic operator in the parabolic basis
we use continued fractions. For the minimal solution of the
recurrence relations we find for $s^{-1} \ll 1$
\bea
\label{SCH-V1-EL216}
\frac{A_{s+1}}{A_{s}} \sim \frac{D^2\omega}{4s}\left (1+
O(\frac{1}{\sqrt{s}})\right).
\eea
Thus we have
$$
A_{s} \sim \frac{\left(\frac{D^2\omega}{4}\right)^s}{s!},
$$
and
\bea
\label{SCH-V1-EL218}
U(\cos\zeta) \sim \sum
\frac{\left(\frac{D^2\omega}{4}\right)^k}{k!}
\cos^{2k}\zeta
\sim
\exp(\frac{D^2\omega}{8}\cos 2\zeta).
\eea
Therefore we see that for this case the function $Z(\cos\zeta; D^2)$
as $\zeta \to i\infty$ is not normalizable.
There is a linearly independent solution of the recurrence relations,
but the coefficients grow even faster.
Hence it follows that the series (\ref{SCH-V1-EL24}) should be truncated.
The condition for series (\ref{SCH-V1-EL24}) be truncated give us already
known formula for energy spectrum (\ref{EL-EN1}) and reduce to the
polynomials:
\bea
\label{SCH-V1-EL219}
U_n^{(\pm k_1, \pm k_2)} (t;  D^2) =
\sum_{s=0}^{n} A_s^{(\pm k_1, \pm k_2)} (D^2) t^{s},
\eea
where now the coefficients
$A_{s} \equiv A_{s}^{(\pm k_1, \pm k_2)} (D^2)$
satisfy the following three-term recurrent relations
\bea
\label{SCH-V1-EL220}
(s+1)(s+1 \pm k_1) A_{s+1}  + \beta_s A_{s}
- \frac{D^2\omega}{4}(n-s+1) A_{s-1} = 0,
\qquad
s=0,1,...n
\eea
with
\bea
\label{SCH-V1-EL221}
\beta_s =
&-&
\frac14 \biggl[(2s+1\pm k_1\pm k_2)^2 +
\frac{D^2 \omega}{2}(2s - n + 4 \pm 4k_1)
\nonumber
\\[2mm]
&-&
\frac{D^2 \omega}{4}(2\pm k_1 \pm k_2)
- \frac{D^4\omega^2}{64}+ \lambda (D^2)\biggr]
\eea
and $A_{-1} = A_{n+1} = 0$.

The recurrent relations (\ref{SCH-V1-EL220}) becomes a system
of $(n+1)$ linear homogeneous equations for the coefficients $A_s$.
Equating the corresponding determinat to zero
\bea
\label{SCH-V1-EL222}
D_n (\lambda) =
\left|
\begin{array}{ccccccc}
\beta_0&(1\pm k_1)&&\cdot&& \\
-\frac{D^2\omega}{4} n&\beta_1&2(2\pm k_1)&\cdot&&\\
\cdot&\cdot&\cdot&\cdot&&\cdot&\cdot\\
&&&\cdot&-\frac{D^2\omega}{2}&\beta_{n-1}&n(n\pm k_1)\\
&&&\cdot&&-\frac{D^2\omega}{4}&\beta_n
\end{array}
\right| = 0
\eea
leads to the algebraic equation of a $(n+1)$ degree which determine
the eigenvalues of elliptic separation constant
$\lambda_{nq}^{(\pm k_1, \pm k_2)}(D^2)$.
The quantum number $q=0,1,2,..n$ enumerate $(n+1)$ roots of eq.
(\ref{SCH-V1-EL222}) and therefore the degree of degeneracy as
in the polar and Cartesian cases for $n$-th energy state equal $n+1$.
It is also known at the corresponding numeration the quantum number
$q$ define the numbers of zeros of polynomial (\ref{SCH-V1-EL219}),
which has exactly $n$ - zeros situated in the open interval
$0 < t < \infty$, and therefore, the elliptic separation constant
$\lambda_{nq}^{(\pm k_1, \pm k_2)}(D^2)$
may be ordered also by the numbers of the nodes
of the eigenfunction of equation (\ref{SCH-V1-EL5}).

\vspace{0.4cm}
\noindent
{\bf 3.3.5. Wave functions}.

\noindent
Thus the the condition of finitness of the solution of eq.
(\ref{SCH-V2-EL5}) allows the following polynomials:
\bea
\label{SCH-V1-EL223}
{\cal I}_{n q}^{(\pm k_1, \pm k_2)} (\zeta;  D^2) =
(\sin\zeta)^{\frac12\pm k_2} \ (\cos\zeta)^{\frac12\pm k_1}
\sum_{s=0}^{n} A_s^{(\pm k_1, \pm k_2)} (D^2) (\cos\zeta)^{2s},
\eea
while the corresponding solution of eq. (\ref{SCH-V1-EL5}) is
\bea
\label{SCH-V1-EL224}
{\cal Z}_{n q}^{(\pm k_1, \pm k_2)} (\zeta;  D^2) =
e^{- \frac{D^2\omega}{16} \cos 2\zeta}
\,\,
{\cal I}_{n q}^{(\pm k_1, \pm k_2)} (\zeta;  D^2)
\eea
We will term the polynomials
${\cal I}_{n q}^{(\pm k_1, \pm k_2)} (\zeta;  D^2)$ as the
{\it associated Ince polynomials}. In the case of $k_1 = k_2 = 1/2$
these polynomials transforms to the four type of the ordinary
{\it Ince polynomials}, which are even or odd with respect to the changes
$\zeta \to -\zeta$ and $\zeta \to \zeta + \pi$ \cite{ARS,MPSTAN}.

At $\zeta = \mu$ the wave functions (\ref{SCH-V1-EL224}) give us the
solution of angular equation (\ref{SCH-V1-EL4}), and for $\zeta = i\nu$
the solution of radial equation (\ref{SCH-V1-EL3}).
For the each of wave functions, radial or angular, corresponds definite
numbers of zeros which could be presented by two quantum numbers $q_1$
and $q_2$, obeying the condition $q_1+q_2=n$. Then the complete elliptic
wave function (\ref{SCH-V1-EL2}) may be written as
\bea
\label{SCH-V1-EL225}
\Psi_{n q_1 q_2}^{(\pm k_1, \pm k_2)} (\nu, \mu; D^2) =
C_{n q_1 q_2} (\pm k_1, \pm k_2; D^2) \,
{\cal Z}_{n q_1}^{(\pm k_1, \pm k_2)} (\mu;  D^2)
\,\,
{\cal Z}_{n q_2}^{(\pm k_1, \pm k_2)} (i\nu;  D^2),
\eea
where $C_{n q_1 q_2} (\pm k_1, \pm k_2; D^2)$ is the normalization
constant. It could be calculated from the condition
\bea
\label{SCH-V1-EL226}
\frac{D^2}{4} \int_{0}^{\infty} d\nu \int_{0}^{\frac{\pi}{2}}
d\mu \, (\cosh^2\nu - \cos^2\mu) \,
\Psi_{n q_1 q_2}^{(\pm k_1, \pm k_2) *} (\nu, \mu; D^2)
\Psi_{n q_1 q_2}^{(\pm k_1, \pm k_2)} (\nu, \mu; D^2)
= \frac{1}{4}
\eea
In same cases it is more convenient to have alternative form for the
wave functions (\ref{SCH-V1-EL225}). Equation (\ref{SCH-V1-EL5})
represents the eigenvalue problem for the separation constant
$\lambda(D^2)$ and the corresponding eigenfanction. It easy to see
that the operator in the left-hand side of eq. (\ref{SCH-V1-EL5})
is invariant under the simulteneous transformations
\bea
\label{SCH-V1-EL226-1}
\zeta \longrightarrow  \zeta + \frac{\pi}{2},
\qquad
D^2 \longrightarrow - D^2,
\qquad
k_1 \longleftrightarrow k_2
\eea
and therefore the sabstitution $D^2 \longrightarrow - D^2$ and
$k_1 \longleftrightarrow k_2$ does not change the set of eigenvalues
of the separation constant $\lambda(D^2)$, but change only their
numeration:
\bea
\label{SCH-V1-EL226-2}
\lambda_{nq_1}^{(\pm k_1, \pm k_2)}(D^2) =
\lambda_{nq_2}^{(\pm k_2, \pm k_1)}(-D^2).
\eea
As a result transformation (\ref{SCH-V1-EL226-1}) transform the
solutions (\ref{SCH-V1-EL224}) into each other
\bea
\label{SCH-V1-EL226-3}
{\cal Z}_{n q_1}^{(\pm k_1, \pm k_2)} (\zeta;  D^2)
\longrightarrow
{\cal Z}_{n q_2}^{(\pm k_2, \pm k_1)} (\zeta + \frac{\pi}{2}; - D^2)
\eea
The coefficients $A_{s}^{(\pm k_2, \pm k_1)} (-D^2)$ are calculated
from the three-term recurrent relations (\ref{SCH-V1-EL220}) after
the changing $D^2 \longrightarrow - D^2$ and $k_1 \longleftrightarrow k_2$.
From (\ref{SCH-V1-EL226-3}) we conclude that elliptic basis
(\ref{SCH-V1-EL225}) can be represented in the form
\bea
\label{SCH-V1-EL226-4}
\Psi_{n q_1 q_2}^{(\pm k_1, \pm k_2)} (\nu, \mu; D^2) =
{\tilde C}_{n q_1 q_2} (\pm k_1, \pm k_2; D^2) \,
{\cal Z}_{n q_1}^{(\pm k_1, \pm k_2)} (\mu;  D^2)
\,\,
{\cal Z}_{n q_2}^{(\pm k_2, \pm k_1)} (i\nu+ \frac{\pi}{2}; - D^2),
\eea
where the quantum number $q_1, q_2$ ($q_1+q_2=n$) again have the meaning
of zeros of the angular and radial wave functions and the constant
${\tilde C}_{n q_1 q_2} (\pm k_1, \pm k_2; D^2)$ satisfy the normalization
condition (\ref{SCH-V1-EL226}).

\vspace{0.4cm}
\noindent
{\bf 3.3.6. Orthogonality relations}.

\noindent

The wave function (\ref{SCH-V1-EL225}) as eigenfunction of the
Hamiltonians are orthogonal $n\not= n'$
\bea
\label{SCH-V1-EL227}
\int_{0}^{\infty} \int_{0}^{\frac{\pi}{2}}
\Psi_{n' q_1 q_2}^{(\pm k_1, \pm k_2) *} (\nu, \mu; D^2)
\Psi_{n q_1 q_2}^{(\pm k_1, \pm k_2)} (\nu, \mu; D^2)
dV = 0
\eea
Eqs. (\ref{SCH-V1-EL3}) and (\ref{SCH-V1-EL4}) allow to prove
the property of {\it double orthogonality} for wave functions
${\cal Z}_{n q}^{(\pm k_1, \pm k_2)}(\zeta; D^2)$, namely
\bea
\label{SCH-V1-EL228}
\int_{0}^{\infty}
{\cal Z}_{n q_2'}^{(\pm k_1, \pm k_2) *} (i\nu; D^2)
{\cal Z}_{n q_2}^{(\pm k_1, \pm k_2)} (i\nu; D^2)
d\nu = 0
\\[2mm]
\label{SCH-V1-EL229}
\int_{0}^{\frac{\pi}{2}}
{\cal Z}_{n q_1'}^{(\pm k_1, \pm k_2) *} (\mu; D^2)
{\cal Z}_{n q_1}^{(\pm k_1, \pm k_2)} (\mu; D^2)
d\mu = 0
\eea
for $q_1\not= q_1'$ and $q_2\not= q_2'$, and therefore when
$q\not= q'$
\bea
\label{SCH-V1-EL230}
\int_{0}^{\infty} d\nu \int_{0}^{\frac{\pi}{2}}
d\mu \, (\cosh^2\nu - \cos^2\mu) \,
\Psi_{n q_1' q_2'}^{(\pm k_1, \pm k_2) *} (\nu, \mu; D^2)
\Psi_{n q_1 q_2}^{(\pm k_1, \pm k_2)} (\nu, \mu; D^2)
= 0
\eea

\vspace{0.4cm}
\noindent
{\bf 3.3.7. Limit $D \to 0$}.

\noindent
Now we will show that in the limit $D \to 0$ the elliptic bases
transforms
to polar one. At the limit $D\to 0$ in eq. (\ref{SCH-V1-EL222})
it is possible to neglect the terms depending of $D^2$, then
\bea
\label{SCH-V1-EL231}
D_n (\lambda) = \beta_0 \beta_1 ...... \beta_n = 0
\eea
Let one of the $\beta_s$ at $s=m$ equal zero: $\beta_m=0$.
Then $\lambda(0) = - (2m+1\pm k_1 \pm k_2)^2$ and therefore at the
limit $D \to 0$ the quantum numbers $q_1$ and $q_2$ transforms to
$m$ and $n-m=n_r$ correspondingly. For $s \not= m$, we have
\bea
\label{SCH-V1-EL232}
\beta_s (D^2) \longrightarrow \beta_s (0) = - (s-m)(s+m+1 \pm k_1 \pm k_2)
\eea
It follows from (\ref{SCH-V1-EL232}) that the three term recurrent
relation (\ref{SCH-V1-EL220}) at $D\to 0$ split to two term recurrent
relations
\bea
\label{SCH-V1-EL233}
(s+1)(s+ 1 \pm k_1 \pm k_2) A_{s+1} + \beta_s A_s = 0,
\qquad
s=0,1,.... m-1
\eea
and
\bea
\label{SCH-V1-EL234}
\beta_s A_s  - \frac{D^2\omega}{4} (n-s+1) A_{s-1} = 0,
\qquad
s= m+1, m+2, ... n
\eea
From two-term recurrent relations (\ref{SCH-V1-EL233}) and
(\ref{SCH-V1-EL233}), we get
\bea
\label{SCH-V1-EL235}
A_{s} (D^2)
\stackrel{D \to 0}{\longrightarrow}
\frac{(-m)_s (m+1 \pm k_1 \pm k_2)_s}{s! (1\pm k_1)_s},
\eea
for $s=0,1,.... m-1$ and
\bea
\label{SCH-V1-EL236}
A_{m+s} (D^2)
\stackrel{D \to 0}{\longrightarrow}
(-1)^m  \,
\frac{(m+1 \pm k_1 \pm k_2)_m \, (m-n)_s }
{s! (2m +2 \pm k_1 \pm k_2)_s \, (1 \pm k_1)_m }
\,
\left(\frac{D^2 \omega}{4}\right)^s,
\eea
for $s = 1,2, ... n-m$, and $(m)_s = \Gamma(m+s)/\Gamma(m)$.
Then according to eq. (\ref{SCH-V2-EL71})
\bea
\label{SCH-V1-EL237}
{\cal Z}_{n q_1}^{(\pm k_1, \pm k_2)} (\mu;  D^2)
&\stackrel{D \to 0}{\longrightarrow}&
(\cos\phi)^{\frac12 \pm k_1}\, (\sin\phi)^{\frac12 \pm k_2}
\sum_{s=0}^{n}\, A_{s} (D^2)  \, (\cos\phi)^{2s},
\\[2mm]
\label{SCH-V1-EL238}
{\cal Z}_{n q_2}^{(\pm k_1, \pm k_2)} (i\nu;  D^2)
&\stackrel{D \to 0}{\longrightarrow}&
(i)^{(\frac12 \pm k_2)}\, e^{-\frac{\omega r^2}{2}}\,
\left(\frac{2r}{D}\right)^{1 \pm k_1 \pm k_2}\,
\sum_{s=0}^{n}\, \frac{A_{s} (D^2)}{D^{2s}} \, (2r)^{2s},
\eea
Substituting here the eqs. (\ref{SCH-V1-EL235}) and (\ref{SCH-V1-EL236}),
we get
\bea
\label{SCH-V1-EL239}
{\cal Z}_{n q_1}^{(\pm k_1, \pm k_2)} (\mu;  D^2)
&\stackrel{D \to 0}{\longrightarrow}&
(\cos\phi)^{\frac12 \pm k_1}\, (\sin\phi)^{\frac12 \pm k_2}
\,
_2F_1 (-m, m+1 \pm k_1 \pm k_2; 1 \pm k_1; \cos^2\phi)
\nonumber\\[2mm]
&=&
\frac{(-1)^m\, m!}{(1 \pm k_1)_m} \,
(\cos\phi)^{\frac12 \pm k_1}\, (\sin\phi)^{\frac12 \pm k_2}
\,
P^{(\pm k_2, \pm k_1)}_m(\phi)
\\[3mm]
\label{SCH-V1-EL240}
{\cal Z}_{n q_2}^{(\pm k_1, \pm k_2)} (i\nu;  D^2)
&\stackrel{D \to 0}{\longrightarrow}&
(i)^{(2m+ \frac12 \pm k_2)}\,
\frac{(m+1 \pm k_1 \pm k_2)_m}
{(1 \pm k_1)_m }
\nonumber\\[2mm]
&\times&
e^{-\frac{\omega r^2}{2}}\,
\left(\frac{2r}{D}\right)^{2m+1 \pm k_1 \pm k_2}\,
_1F_1 (-n_r; 2m+2 \pm k_1 \pm k_2; \omega r^2)
\nonumber\\[2mm]
&=&
(i)^{(2m+ \frac12 \pm k_2)}\,
\frac{n_r!\, (m+1 \pm k_1 \pm k_2)_m}
{(1 \pm k_1)_m (2m + 1 \pm k_1 \pm k_2)_m}\,
\nonumber\\[2mm]
&\times&
e^{-\frac{\omega r^2}{2}}\,
\left(\frac{2r}{D}\right)^{2m+1 \pm k_1 \pm k_2}\,
L_{n_r}^{2m+1 \pm k_1 \pm k_2} (\omega r^2)
\eea
From these formulas and normalization condition (\ref{SCH-V1-EL226})
we finally get that te elliptic basis (\ref{SCH-V1-EL225}) at the
limit $D\to 0$ becomes the polar basis
\bea
\label{SCH-V1-EL241}
\Psi_{n q_1 q_2}^{(\pm k_1, \pm k_2)} (\nu, \mu; D^2)
\stackrel{D \to 0}{\longrightarrow}
\Psi_{n_r m}^{(\pm k_1, \pm k_2)} (r, \phi)
\eea

\vspace{0.4cm}
\noindent
{\bf 3.3.8. Limit $D \to \infty$}.

\noindent
Let us now to investigate the Cartesian limit. At $D\longrightarrow \infty$
we can neglect all final terms and again the determinat (\ref{SCH-V1-EL222})
transforms to product of diagonal elements. Let now one of the $\beta_s$
for  $s=n_1$ equal zero: $\beta_{n_1}=0$, then
\bea
\label{SCH-V1-EL242}
\lambda_{n q_1}^{(\pm k_1, \pm k_2)} (D^2)
\stackrel{D \to \infty}{\longrightarrow}
\lambda_{n n_1}^{(\pm k_1, \pm k_2)} (D^2)
=
\frac{D^4 \omega^2}{64}
- \frac{D^2 \omega}{4} (4n_1 - 2n + 6 \pm 7 k_1 \mp k_2)
\eea
here $n_1 = 0,1,2,...$. Analogically it is easy obtain
\bea
\label{SCH-V1-EL243}
\lambda_{n q_2}^{(\pm k_2, \pm k_1)} (-D^2)
\stackrel{D \to \infty}{\longrightarrow}
\lambda_{n n_1}^{(\pm k_2, \pm k_1)} (-D^2)
=
\frac{D^4 \omega^2}{64}
+
\frac{D^2 \omega}{4} (4n_2 - 2n + 6 \pm 7 k_2 \mp k_1)
\eea
with $n_2=0,1,2,...$, and from the condition (\ref{SCH-V1-EL226-2})
follows that $n_1+n_2=n$ and $n_1, n_2$ are the Cartesian quantum
number. From the eqs. (\ref{SCH-V1-EL242}) and (\ref{SCH-V1-EL242}),
and (\ref{SCH-V1-EL221}) it is easy to get ($s \not= n_1$, and
$s \not= n_2$)
\bea
\label{SCH-V1-EL244}
\beta_{s} (D^2)
\stackrel{D \to \infty}{\longrightarrow}
D^2 \omega (s-n_1),
\qquad
\beta_{s} (-D^2)
\stackrel{D \to \infty}{\longrightarrow}
-D^2 \omega (s-n_2).
\eea
These formulas and the condition $A_{-1}=0$ shows that in the limit
$D\to \infty$ the three-term recurrent relation (\ref{SCH-V1-EL220})
transforms to two-term recurrent relations
\bea
\label{SCH-V1-EL245}
(s+1)(s+1\pm k_1) A_{s+1}(D^2) - \frac{D^2\omega}{4} (s-n_1) A_{s}(D^2) = 0,
\qquad
0 \leq s \leq n_1-1
\\[2mm]
\label{SCH-V1-EL245-0}
(s+1)(s+1\pm k_2) A_{s+1}(-D^2) + \frac{D^2\omega}{4} (s-n_2) A_{s}(-D^2)
= 0,
\qquad
0 \leq s \leq n_2-1
\eea
Analogically using the condition $A_{n+1}=0$, we obtain
\bea
\label{SCH-V1-EL246}
(s-n_1) A_{s}(D^2) + \frac{1}{4} (n-s+1) A_{s-1}(D^2)
= 0,
\qquad
n_1+1 \leq s \leq n
\\[2mm]
\label{SCH-V1-EL246-0}
(s-n_2) A_{s}(-D^2) + \frac{1}{4} (n-s+1) A_{s-1}(-D^2)
= 0,
\qquad
n_2+1 \leq s \leq n
\eea
From the two-term recurrent relations (\ref{SCH-V1-EL245}) -
(\ref{SCH-V1-EL245-0}) we have
\bea
\label{SCH-V1-EL247}
A_{s}(D^2)
&\stackrel{D \to \infty}{\longrightarrow}&
\frac{(D^2 \omega)^s}{4^s} \,
\frac{(-n_1)_s}{s!\, (1\pm k_1)_s},
\qquad
0 \leq s \leq n_1
\\[2mm]
\label{SCH-V1-EL247-0}
A_{n_1+s}(D^2)
&\stackrel{D \to \infty}{\longrightarrow}&
\frac{A_{n_1}(D^2)}{4^s}\,
\frac{(n_1-n)_s}{s!},
\qquad
1 \leq s \leq n - n_1
\eea
and the same for $A_{s}(-D^2)$
\bea
\label{SCH-V1-EL248}
A_{s}(-D^2)
&\stackrel{D \to \infty}{\longrightarrow}&
\frac{(-D^2 \omega)^s}{4^s} \,
\frac{(-n_2)_s}{s!\, (1\pm k_2)_s},
\qquad
0 \leq s \leq n_2
\\[2mm]
\label{SCH-V1-EL248-0}
A_{n_2+s}(-D^2)
&\stackrel{D \to \infty}{\longrightarrow}&
\frac{A_{n_2}(-D^2)}{4^s}\,
\frac{(n_2-n)_s}{s!},
\qquad
1 \leq s \leq n - n_2
\eea
We can investigate now the limit $D\to \infty$ in the wave functions.
Taking into account the conditions (\ref{SCH-V2-EL710}) and formulas
(\ref{SCH-V1-EL247})-(\ref{SCH-V1-EL248-0}), we obtain the following
results
\bea
\label{SCH-V1-EL249}
{\cal I}_{n q_1}^{(\pm k_1, \pm k_2)} (\mu;  D^2)
&\stackrel{D \to \infty}{\longrightarrow}&
\left(\frac{2x}{D}\right)^{\frac12\pm k_1}\,
\sum_{s=0}^{n} \frac{A_s^{(\pm k_1, \pm k_2)} (D^2)}{(D^2)^{s}}
(4x^2)^{s}
\nonumber\\[2mm]
&=&
\left(\frac{2x}{D}\right)^{\frac12\pm k_1}\,
{_1F_1} (-n_1; \, 1 \pm k_1; \, \omega x^2),
\\[2mm]
{\cal I}_{n q_2}^{(\pm k_2, \pm k_1)} (i\nu+\frac{\pi}{2}; -D^2)
&\stackrel{D \to \infty}{\longrightarrow}&
\left(\frac{2iy}{D}\right)^{\frac12\pm k_2}\,
\sum_{s=0}^{n} \frac{A_s^{(\pm k_2, \pm k_1)} (-D^2)}{(-D^2)^{s}}
(4y^2)^{s}
\nonumber\\[2mm]
&=&
\left(\frac{2iy}{D}\right)^{\frac12\pm k_2}\,
{_1F_1} (-n_2; \, 1 \pm k_2; \, \omega y^2).
\eea
Using the connection between hypergeometrical functions and Lagerre
polynomials \cite{BE}
\bea
\label{SCH-V1-EL250}
L^{\alpha}_n (x) = \frac{(\alpha+1)_n}{n!}
{_1F_1} (-n; \, 1 + \alpha; \, x ).
\eea
finally we get
\bea
\label{SCH-V1-EL251}
\Psi_{n q_1 q_2}^{(\pm k_1, \pm k_2)} (\nu, \mu; D^2)
&\stackrel{D \to 0}{\longrightarrow}&
\frac{{\tilde C}_{n q_1 q_2} (\pm k_1, \pm k_2; D^2)}
{D^{1\pm k_1 \pm k_2}}\,
e^{ - \frac{\omega}{2} (x^2 + y^2)}
\nonumber\\[2mm]
&\times&
(2x)^{\frac12\pm k_1}\,
(2y)^{\frac12\pm k_2}\,
L^{\pm k_1}_{n_1} (\omega x^2)
L^{\pm k_2}_{n_2} (\omega y^2)
\eea
which after taking the limit in the normalization constant coinside
with the Cartesian basis of singular oscillator.

\section{Conclusions and summary}

Let us here to summarize the presented investigation of superintegrable
potentials $V_1$ and $V_2$.

We have determined that solution of Schr\"odinger equation for the
potentials $V_1$ maybe constracted via separation of variables with
two different ways. One of them is exploting the separation of variables
in Cartesian coordinates, which lead two independent exactly-solvable
equations (\ref{V1-SHEQ0}) and (\ref{V1-SHEQ1}), the each of them
represent the one-dimensional {\it non-parametric} spectral problem
where the cartesian separation constants $\lambda_i$ play the role of
energy. It admit to get the solution in form of Lagerre and Hermite
polynomials, quantize both separation constants and as result to obtain
energy spectrum for two-dimensional Schr\"odinger equation.
The second way is the separation of variables in parabolic coordinate.
We have shown that the separation procedure reduce to {\it one-parametric}
(parabolic separation constant) Schr\"odinger type differential
equation (\ref{V1-EQ202}), living in the complex plane.
It have been proven that the requirement of convergence for
solutions of eq. (\ref{V1-EQ202}) at the singular points
$\mu = \pm \infty$ and $\mu= i\infty$ lead only to polynomial solutions
(\ref{V1-EQ20}) with restriction for energy spectrum $E$
in form (\ref{V2-E1}) and at the fixed energy (or quantum number $n$)
give the spectrum of separation constant as root of nth-degree
algebraic equation.
In difference of the previous case, the power expansions of polynomial
solutions satisfy three-term recurrence relations and cannot be rewritting
in explicit form and admit only numerical description. For this reason
no sence to term the equation (\ref{V1-EQ202}) as exactly-solvable.

At the other hand side, the substitution of the formula of energy
spectrum in the eq. (\ref{V1-EQ202}) bring us to equation
\be
\label{V1-EQ3}
\left[-\frac{d^2}{d\mu^2} + \left(\omega^2\mu^6 + k_1 \mu^4
+ \left [\frac{k_1^2}{4\omega^2} - \omega(4n + 4 \pm 2k_2)
\right] \mu^2 + \frac{k_2^2 -\frac{1}{4}}{\mu^2}\right)\right]
Z_n (\mu) = \lambda Z_n (\mu),
\ee
which on the real axis completely coincides for $k_1 = 4\beta \omega^2$
and $1 \pm k_2 = 2\delta$, with the one-dimensional spectral problem
(\ref{SCH-V0-ANHAR0}), and term as quasi-exactly solvable problem.
Now it is easy to understand the birth mechanism of quasi-exactly
solvable systems. The requirement of convergence just in real space
(which is possible to determin following the \cite{VINET} as the
dimensional reduction) in singular points $\mu = \pm \infty$ performe
together with the polynomial solutions in form (\ref{V1-EQ0-8})
(with restriction on real or imagenery axis), which is already non
complete, also non polynomial part.
We also can shed a light on the mistery of zeros of polynomial
$P_n(x^2)$. Indeed, the substitution of the wave function
(\ref{SCH-POLYNOM1}) in Schr\"odinger equation with potential
(\ref{SCH-V0-ANHAR0}) lead to the differential equation for
polynomial $P_n(x^2)$ in the same form as equation (\ref{V1-EQ7-1})
(in variable $x^2 = z$), but with difference that the physical region
of eq. (\ref{V1-EQ7-1}) is whole real axis $z\in (-\infty, \infty)$,
and therefore all zeros (for positive and negative $x^2$) of $P_n(x^2)$
corresponds to the zeros of two-dimensional eigenfunction of
singular anisotropic oscillator in parabolic coordinates.

We also shown the situation have repeated in the case of second
potential (\ref{V2}). We have determined that the separation
of variables in two-dimensional elliptic coordinates lead to Schr\"odinger
type equation (\ref{SCH-V1-EL5}) in complex plane and requirement of
convergence at the point $\zeta = 0, 2\pi$ and $\zeta=i\infty$ deduce
to polynomial solutions and difine the energy spectrum (\ref{EL-EN1}).
So, it generate trigonometric and hyperbolic quasi-exactly solvable
systems (see potentials 5 and 8 in \cite{USHV1}) in form of
\bea
\frac{d^2 X}{d\nu^2}
&+&
\left[\left(\frac{\alpha^2}{4} + \alpha (2n+2\pm k_1 \pm k_2)\right)
\cosh^2\nu - \frac{\alpha^2}{4} \cosh^4\nu
- \frac{k^2_1 - \frac{1}{4}}{\sinh^2\nu} +
\frac{k^2_2 - \frac{1}{4}}{\cosh^2\nu} + \lambda
\right]X = 0,
\nonumber
\\[3mm]
\frac{d^2 Y}{d\mu^2}
&-&
\left[\left(\frac{\alpha^2}{4} + \alpha (2n+2\pm k_1 \pm k_2)\right)
\cos^2\mu - \frac{\alpha^2}{4}\cos^4\mu +
\frac{k^2_1 - \frac{1}{4}}{\cos^2\mu} +
\frac{k^2_2 - \frac{1}{4}}{\sin^2\mu} + \lambda
\right] Y = 0,
\nonumber
\eea
where $\alpha= D^2\omega/2$.
Thus we have proven that in the base of quasi-exactly solvability
phenomena stay the dimensional reduction of superintegrable
systems on the one-dimension
\footnote{Really we can express our observation in the form of
following hypothesis: \it all quantum-mechanical problems which
known for today as one-dimensional quasi-exactly solvable systems
could be determine via separation of variables in N-dimensional
Schr\"odinger equation for superintegrable systems.}.

This analogy prompt us to use the term of quasi-exactly solvability
for the equations of type (\ref{V1-EQ202}) or (\ref{SCH-V1-EL5}),
define in the complex plane and which are not exactly-solvable but
also admit complete set of polynomial solution. Thus we suggest to
call the quantum mechanical systems {\bf first-order quasi-exactly
solvable} if the polynomial solution one-parametric
differential equation of the kind of Schr\"odinger equation
or $N$ -dimensional equation after separation of variables is defined
through three-term recurrence relations and the discrete eigenvalues
could be calculated only numerically as the solutions of algebraic
equations. Accordingly this definition systems
(\ref{V1-EQ202}) and (\ref{SCH-V1-EL5}) are  first order quasi-exactly
solvable. What about quasi-exactly solvable systems in his old sense,
it more convenient to call as {\it partially quasi-exactly solvable},
because only part of solution is polynomial.

It is easy to generelize our definition for $N$-dimensional
quasi-exactly solvable systems. Let us consider the Helmholtz equation
on $N$-dimensional sphere
$S_N$: $u_1^2 + u_2^2+... u_{N+1}^2 = R^2$
\bea
\label{HELM-HAM1}
\frac{1}{2} \Delta_{LB} \Psi + E \Psi = 0,
\eea
where the Laplace-Beltrami operator is
\bea
\label{HELM-HAM2}
\Delta_{LB} =  - \frac{1}{2R^2} \sum_{i,k=1}^{N+1} L_{ik}^2,
\qquad
L_{ik} = - i\left(u_{i}\frac{\partial}{\partial u_k}
-u_{k}\frac{\partial}{\partial u_i}\right),
\eea
The separation of variables in $N$-dimensional ellipsoidal coordinates
\cite{KALM1,KALM2,KALM3}
\bea
\label{HELM-HAM3}
u_i^2 = \frac{\Pi_{j=1}^N (\rho_j-a_i)}
{\Pi_{j\not=i}^{N+1}(a_j-a_i)}
\qquad
i = 1,2,...N+1
\eea
with
\bea
\label{HELM-HAM4}
a_1 < \rho_1 < a_2 < \rho_2 < a_3 < ...... < \rho_N < a_{N+1},
\eea
lead to system of $N$ - equivalent differential equations each in the
region $a_{i}< \rho_i< a_{i+1}$. They have the following form
\bea
\label{HELM-HAM5}
4 \,
\sqrt{\Pi_{i=1}^{N+1} (\rho-a_i)}\, \frac{\partial}{\partial \rho} \,
\sqrt{\Pi_{i=1}^{N+1} (\rho-a_i)}\, \frac{\partial\psi}{\partial \rho}
+
\left( E \rho^{N-1} +
\sum_{j=2}^{N}\lambda_j \rho^{N-j}\right)\psi = 0
\eea
and together with the energy $E$ contain the $(N-1)$ separation
constants $\lambda_i$ $(i=1,2,...N-1)$ as parameter. This is the case
of complete nonseparation of separation constants. Equation
(\ref{HELM-HAM5}) is the generalized Lame' equation and falls
into a class of equations of the Fuchsian type with
$N+2$ singularities in the points:
$\{a_{1}, a_{2}, ..... a_{N}, a_{N+1}, \infty \}$
\footnote{$(a_{1}, a_{2}, .... a_{N}, a_{N+1})$ are elementary
singularities with indices (0,1/2 ) and a point at infinity is regular.}.

Consider $N=2$ when eq. (\ref{HELM-HAM5}) reduce to the famous Lam'e
equation. It is possible to prove that condition of finiteness for
eigenfunctions $\psi$ in the whole region for variable $\rho$
including the singular point $(a_1, a_2, a_3)$ perform only the
polynomial solutions (Lam'e polynomials) which provide the quatization
of energy in form $E=\ell(\ell+1)/R^2$, where $\ell = 0,1,2...$ and
simultaneously determine the spectrum of separation constant
$\lambda_i(R)$ as a solution of algebraic equation. It is well known
that Lam'e polynomials can be represented by the series expansion around
one of the singularities with the three-term recurrence relation
for corresponding coefficients. Accordingly of our definition the
equation (\ref{HELM-HAM5}) at $N=2$ in interval $(a_1, a_3)$
give as the example of first order quasi-exactly solvable system.
It is also obviously that reduction into intervals $(a_1, a_2)$
or $(a_2, a_3)$ for the fixed values of $E=\ell(\ell+1)/R^2$
admit both polynomial and non-polynomial solutions. The system become
partially quasi-exactly solvable.

Let us now consider the next dimension $N=3$ which is more complicated.
For $N=3$ equation (\ref{HELM-HAM5}) already contain together energy
parameter $E$ also two separation constants $\lambda_1$ and $\lambda_2$.
As in previous case it is also possible to prove that the requirement
of finiteness for wave function simultaneously in three intervals
$(a_1, a_2)$, $(a_2, a_3)$ and $(a_3, a_4)$ lead to polynomial
solutions and provide quantization of energy according the formula
$E=J(J+2)/R^2$. Accordingly to \cite{AKPS} the generalized Lam'e
polynomial is derived as expansion around one of the singularities
$a_i$ ($i=1,2,3,4$) of the equation (\ref{HELM-HAM5}) where the
coefficients of this expansion obey the four-term recurrence relations.
The obtained from this recurrence relations system of homogeneous
algebraic equations is overcomplete since the number of equations
is larger than the number of unknowns (expansions coefficients),
and the corresponding matrix is rectangular. Concerning a homogeneous
systems of this type, it is known that a necessary and sufficient
condition for the existence of a nontrivial solution is equality to
zero of all determinants, however, as it is proved in \cite{AKPS},
for our system it is sufficient that only two determinants, resulting
from this system by eliminating the last and the next to last rows,
be equal to zero. Therefore we are coming to the system of two
algebraic equations which give a quatization of two separation
constant $\lambda_1(R)$ and $\lambda_2(R)$. So, we term the
equation (\ref{HELM-HAM5}) for $N=3$ in the whole interval of
variable $\rho\in (a_1, a_4)$ as
{\it second order quasi-exactly solvable}.

Now we can determine the $N$th order quasi-exactly solvable
systems following way: the quantum mechanics system is
$N$th - {\it order quasi-exactly solvable} if the polynomial
solution of the $N$ parametric differential equation of the
Schr\"odinger equation type (or $N+1$ - dimensional
Schr\"odinger equation after separation of variables) is defined
through $N+1$-term recurrent relations and the spectrum of eigenvalues
of separation constant could be calculated only numerically as the
solutions of algebraic (or systems of algebraic) equations.

Thus in this work we classify four kind of solvability:
{\bf exactly-solvable}, {\bf quasi-exactly solvable},
{\bf partially quasi-exactly solvable}, and {\bf non-exactly solvable}
systems.

The interesting question which we not consider here is following:
is it possible to classify all polynomials which coeeficients in power
series expansions can be determine through three-term or more higher
order recurrence relations as it known for classical polynomials
as Jacoby, Gegenbauer, Legandre, Lagerre and Hermite polynomials?

\section*{Acknowledgements} G.S.P. thank the support of the Direcci\'on
General de Asuntos del Personal Acad\'emico, Universidad Nacional
Aut\'onoma de M\'exico ({\sc dgapa--unam}) by the grant 102603 {\it
Optica Matem\'atica} and also {\sc sep-conacyt} project 44845.

\end{document}